\documentclass[12pt,draftcls,journal,onecolumn]{IEEEtran}

\usepackage{amsmath,amssymb}
\usepackage{graphicx,psfrag}
\usepackage{color}
\usepackage{cite}
\usepackage{subfigure}
\usepackage{url}
\usepackage{array}
\usepackage{enumerate}
\usepackage{amsthm}
\usepackage{comment}
\usepackage{float}
\usepackage{balance}
\usepackage{algorithm}
\usepackage{algorithmic}
\usepackage{balance}
\usepackage{subfigure}
\usepackage{bm}
\usepackage{diagbox}

\newtheorem{theorem}{Theorem}
\newtheorem{lemma}{Lemma}
\newtheorem{proposition}{Proposition}
\newtheorem{corollary}{Corollary}
\newtheorem{remark}{Remark}

\begin{document}

\title{Privacy against a Hypothesis Testing Adversary}

\author{Zuxing Li, \IEEEmembership{Member, IEEE}, Tobias J. Oechtering, \IEEEmembership{Senior Member, IEEE},\\and Deniz G\"{u}nd\"{u}z, \IEEEmembership{Senior Member, IEEE}
\thanks{This work has been supported by the Swedish Research Council (VR) within the CHIST-ERA project COPES under Grant 2015-06815 and within the project CLONE under Grant E0628201, and by the UK Engineering and Physical Sciences Research Council (EPSRC) under Grant EP/N021738/1 within the CHIST-ERA project COPES. This work was presented in part at the 2017 IEEE International Symposium on Information Theory \cite{zuxing2017}.}
\thanks{Z. Li and T. J. Oechtering are with the School of Electrical Engineering and Computer Science, KTH Royal Institute of Technology, Stockholm 100 44, Sweden (e-mail: zuxing@kth.se; oech@kth.se).}
\thanks{D. G\"{u}nd\"{u}z is with the Department of Electrical and Electronic Engineering, Imperial College London, London SW7 2BT, United Kingdom (e-mail: d.gunduz@imperial.ac.uk).}
}


\maketitle

\begin{abstract}
Privacy against an adversary (AD) that tries to detect the underlying privacy-sensitive data distribution is studied. The original data sequence is assumed to come from one of the two known distributions, and the privacy leakage is measured by the probability of error of the binary hypothesis test carried out by the AD. A management unit (MU) is allowed to manipulate the original data sequence in an online fashion, while satisfying an average distortion constraint. The goal of the MU is to maximize the minimal type II probability of error subject to a constraint on the type I probability of error assuming an adversarial Neyman-Pearson test, or to maximize the minimal error probability assuming an adversarial Bayesian test. The asymptotic exponents of the maximum minimal type II probability of error and the maximum minimal error probability are shown to be characterized by a Kullback-Leibler divergence rate and a Chernoff information rate, respectively. Privacy performances of particular management policies, the memoryless hypothesis-aware policy and the hypothesis-unaware policy with memory, are compared. The proposed formulation can also model adversarial example generation with minimal data manipulation to fool classifiers. Lastly, the results are applied to a smart meter privacy problem, where the user's energy consumption is manipulated by adaptively using a renewable energy source in order to hide user's activity from the energy provider.
\end{abstract}



\section{Introduction}
\label{section1}
Developments in information technology have drastically changed people's lives, e.g., smart homes, smart grids, and e-health. Most of these technologies are based on intelligent algorithms that provide better use of limited resources thanks to their ability to gather and process more information about users' behaviors and preferences. This population-scale collection of personal user data in return of various utilities is a growing concern, as the collected data can be misused beyond their intended application. Therefore, guaranteeing user privacy while continuing to deliver the benefits of such technologies is a fundamental research challenge that must be tackled in order to enable the adoption of these technologies without being concerned about privacy risks.

Privacy has been studied extensively in recent years for various information systems considering different privacy measures \cite{dwork2006,monedero2010,calmon2012,borzoo20181,sweeney2002,jesus2015,issa2016}. Differential privacy \cite{dwork2006,acs2011,zhao2014} and mutual information \cite{wyner1975,jesus2015,mhanna2015} are probably the most commonly studied privacy measures. In \cite{farokhi2018}, Fisher information is proposed as a privacy measure, and is shown to be a lower bound on the mean squared error of an unbiased estimation on the privacy. Kullback-Leibler divergence is used as a privacy measure in \cite{zuxing20152,nadendla2016}, while total variation distance is considered in \cite{borzoo20182}, which provides a bound on the privacy leakage measured by mutual information or maximal leakage \cite{issa2016}.

Privacy is particularly difficult to guarantee against ADs employing modern data mining techniques, which are capable of identifying user behavior with large datasets. With such attacks in mind, we propose a novel formulation of the privacy leakage problem that can be applied to many information sharing applications. We assume that the original data is manipulated by the MU and then shared with a remote entity in an online manner. Our goals are to guarantee that the shared data satisfies a certain utility constraint, measured by the average distortion between the original data sequence and the shared sequence, and meanwhile to limit the inference capability of the AD, who might be the legitimate receiver of the shared data. We assume that the AD performs an optimal hypothesis test based on the shared data sequence (or the adversarial observation sequence) and tries to determine a privacy-sensitive underlying hypothesis. Hypothesis test has previously been considered in the privacy context \cite{willenborg2001,zuxing2015,liao2016,liao2018,sreekumar2018}. However, different from these previous works, in which the goal is to increase the reliability of the hypothesis test while guaranteeing privacy, we measure the privacy risk by the error probability of the corresponding adversarial hypothesis testing problem.

Our problem formulation can also be considered as a model for generating adversarial examples to attack classifiers, which is a very popular research area due to the increasing adoption of machine learning techniques in all domains \cite{huang2011,kurakin2016,athalye2017,tramer2017}. These attacks consist of generating examples that can fool a classifier, despite being very similar to the true distribution of the data that has been used to train the classifier. Deep neural networks are particularly vulnerable to such attacks \cite{goodfellow2014}. In our setting, the AD can be considered as the classifier, and we are aiming at generating adversarial observation sequences that are similar to the original data sequence under a given distortion constraint, yet will make the detection of the AD as unreliable as possible.

A similar adversarial signal processing problem is studied in \cite{barni2013,tondi2015,barni2016}, which also considers independent identically distributed (i.i.d.) original data sequences, distortion-constrained data manipulation, and hypothesis test based on manipulated sequences. However, the results in this paper and in \cite{barni2013,tondi2015,barni2016} do not imply each other due to the following fundamental differences: i) We consider an MU to degrade the hypothesis testing accuracy of a passive but informed AD, and formulate the problems as worst-case analyses; while \cite{barni2013,tondi2015,barni2016} consider an active attacker to degrade the hypothesis testing accuracy of a defender and formulate the problem as a zero-sum game; ii) we assume the AD to always perform an optimal hypothesis test without the restriction that the hypothesis test has to be based on the type of the manipulated data sequence as imposed in \cite{barni2013,tondi2015,barni2016}; iii) we also study the scenario in which the MU does not know the hypothesis \textit{a priori}.

We study the privacy performance by focusing on the asymptotic error exponent in two different settings. We consider i) the asymptotic exponent of the maximum minimal type II probability of error assuming an adversarial Neyman-Pearson test setting, and ii) the asymptotic exponent of the maximum minimal error probability assuming an adversarial Bayesian test setting. We show that the asymptotic error exponent can be characterized by a Kullback-Leibler divergence rate in the first setting, and by a Chernoff information rate in the second. In particular, we prove that the asymptotic error exponent achieved by the optimal memoryless hypothesis-aware policy reduces to a single-letter Kullback-Leibler divergence, or a single-letter Chernoff information. We also consider the hypothesis-unaware policies with memory, and show for both settings that, the asymptotic error exponent achieved by the optimal memoryless hypothesis-aware policy is an upper bound on the exponent achieved by the optimal hypothesis-unaware policy with memory. Here, we generalize the privacy problem model we introduced in \cite{zuxing2017}, present the omitted proofs of Lemma 1, Lemma 2, Theorem 3, Theorem 4, and modify the proof of Theorem 1. Additionally, we analyze here the adversarial Bayesian hypothesis test setting.

As an application of the presented theoretical framework, we will consider a smart meter that reports the energy supply data to an energy provider (EP) at regular time intervals. The EP, which is the legitimate receiver of the smart meter data, can be considered as the AD in this setting, that tries to mine the smart meter data beyond the intended utility application, e.g., to infer the privacy-sensitive user behavior information: presence at home, or electrical appliance usage patterns. Various privacy-preserving techniques have been developed for the smart meter privacy problem in recent years, that can be classified into two groups \cite{giaconi20181}: The methods in the first group modify the smart meter readings in order to confuse the EP; while the second group of methods directly modify the energy supply pattern to achieve the same goal \cite{tan2013,varodayan2011,zuxing20152,jesus2015,giaconi2018,acs2011,zhao2014, fan2017}. The advantage of the second approach is that the smart meters report truthful readings; therefore, the EP can reliably use their readings for energy provisioning. Both approaches can be formulated in the general framework proposed in this paper.

In the following, unless otherwise specified, we will denote a random variable by a capital letter, e.g., $X$, its realization by the lower-case letter, e.g., $x$, and its alphabet by the calligraphic letter, e.g., $\mathcal{X}$. Let $X_{t}^{k}$, $x_{t}^{k}$, and $\mathcal{X}_{t}^{k}$ denote a random sequence $(X_{t},\dots,X_{k})$, its realization $(x_{t},\dots,x_{k})$, and its alphabet $\mathcal{X}_{t}\times\dots\times\mathcal{X}_{k}$, respectively. For simplification, $X^{k}$, $x^{k}$, and $\mathcal{X}^{k}$ are used when $t=1$. We use $\textnormal{D}(\cdot||\cdot)$ to denote Kullback-Leibler divergence, $\textnormal{D}_{\tau}(\cdot||\cdot)$ to denote $\tau$-th order R\'{e}nyi divergence, $\textnormal{C}(\cdot,\cdot)$ to denote Chernoff information, and $|\cdot|$ to denote set cardinality.

\begin{figure}
\centering
\includegraphics[scale=0.45]{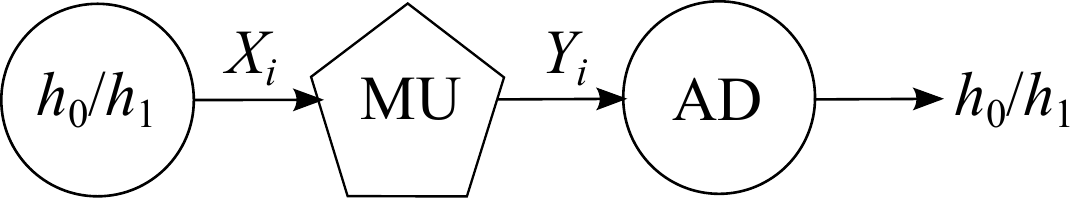}
\caption{The privacy problem model. The original data sequence $\{X_{i}\}$ is i.i.d. generated under the privacy hypothesis $h_{0}$ or $h_{1}$. The management unit (MU) follows a policy to determine the random adversarial observation sequence $\{Y_{i}\}$ that has to satisfy a long-term average distortion constraint. The informed adversary (AD) makes an optimal hypothesis test on the privacy based on the observations.}
\label{figure1}
\end{figure}

\section{System Model}
\label{section2}
The considered privacy problem is illustrated in Fig. \ref{figure1}. The original data sequence $\{X_i\}$ is assumed to come from one of two possible distributions depending on the binary privacy hypothesis $H$, which can be $h_{0}$ or $h_{1}$. Let $p_{0}$ and $p_{1}$ denote the prior probabilities of hypotheses $h_{0}$ and $h_{1}$, respectively, and without loss of generality, we assume $0<p_{0}\leq p_{1}<1$. In the following, we use the notation $\cdot|h_{j}$ for $\cdot|H=h_{j}$, $j\in\{0,1\}$, to denote a random variable under hypothesis $h_{j}$. Under hypothesis $h_{0}$ (resp. $h_{1}$), the data $X_{i}$ at time slot $i$ is i.i.d. generated according to $p_{X|h_{0}}$ (resp. $p_{X|h_{1}}$) where $p_{X|h_{0}}$ and $p_{X|h_{1}}$ are defined on the same finite alphabet $\mathcal{X}$ and satisfy $\mathcal{X}\subset\mathbb{R}$, $-\infty<\min\mathcal{X}<\max\mathcal{X}<\infty$, and $0<\textnormal{D}(p_{X|h_{0}}||p_{X|h_{1}})<\infty$.

At any time slot $i$, the MU follows a management policy $\gamma_{i}$ to determine the (random) adversarial observation $Y_{i}$ based on the data sequence $x^{i}$, the previous adversarial observations $y^{i-1}$, and the true hypothesis $h$ as $Y_{i}=\gamma_{i}(x^{i},y^{i-1},h)$, which can also be represented by the corresponding conditional pmf $p_{Y_{i}|X^{i},Y^{i-1},H}$. Let $\mathcal{Y}$ denote the finite adversarial observation alphabet at any time slot, which satisfies $\mathcal{Y}\subset\mathbb{R}$ and $-\infty<\min\mathcal{Y}\leq\max\mathcal{Y}<\infty$. Let $\gamma^{n}\triangleq\{\gamma_{i}\}_{i=1}^{n}:\mathcal{X}^{n}\times\mathcal{H}\to\mathcal{Y}^{n}$ denote a management policy over an $n$-slot time horizon. The following average distortion constraint is further imposed:
\begin{equation}
\textnormal{E}\left[\left.\frac{1}{n}\sum_{i=1}^{n}d(X_{i},Y_{i})\right|h_{j}\right]\leq s,\textnormal{ }j=0,1,
\label{equation2}
\end{equation}
where $d(\cdot,\cdot):\mathcal{X}\times\mathcal{Y}\to\mathbb{R}_{+}$ is an additive distortion measure. A management policy that satisfies (\ref{equation2}) over an $n$-slot time horizon is denoted by $\gamma^{n}(s)$. The distortion constraint may be imposed either to guarantee a utility requirement from the data, or due to the availability of limited resources to manipulate the data sequence $\{X_i\}$, e.g., smart meter privacy exploiting physical energy sources \cite{giaconi2018}.

We consider that an AD has access to the observations $y^{n}$, and is fully informed about the prior probabilities of the hypotheses, the original data statistics, as well as the adopted management policy, i.e., the AD knows $p_{0}$, $p_{1}$, $p_{X^{n}|h_{0}}$, $p_{X^{n}|h_{1}}$, $\gamma^{n}(s)$, and hence the resulting adversarial observation statistics $p_{Y^{n}|h_{0}}$, $p_{Y^{n}|h_{1}}$. In this work, the informed AD is assumed to make an optimal hypothesis test on the binary privacy hypothesis, and the privacy risk is measured by the probability of error of this adversarial hypothesis test. In the following, we will study the privacy leakage as a Neyman-Pearson hypothesis testing problem and a Bayesian hypothesis testing problem. The corresponding optimal privacy performances will be characterized in the asymptotic regime.

\section{Privacy-Preserving Data Management against a Neyman-Pearson Hypothesis Testing Adversary}
\label{section3}
In this section, the privacy leakage is modeled as a Neyman-Pearson hypothesis test performed by the informed AD. We assume that the AD has a maximum tolerance towards false positives, imposed by a maximum type I probability of error, and its goal is to minimize the type II probability of error under this constraint. Given a management policy $\gamma^{n}(s)$ and the resulting $p_{Y^{n}|h_{0}}$, $p_{Y^{n}|h_{1}}$, we define the minimal type II probability of error of the AD under an upper bound on the type I probability of error as
\begin{equation*}
\beta(n,\varepsilon,\gamma^{n}(s))\triangleq\min_{\mathcal{A}_{n}\subseteq\mathcal{Y}^{n}}\{p_{Y^{n}|h_{1}}(\mathcal{A}_{n})|p_{Y^{n}|h_{0}}(\mathcal{A}_{n}^{c})\leq\varepsilon\},
\end{equation*}
where $\mathcal{A}_{n}$ and $\mathcal{A}_{n}^{c}$ denote the decision regions for $h_{0}$ and $h_{1}$ of the AD, respectively. On the other hand, the privacy-preserving design objective of the MU is to maximize the probability of error of the AD. More specifically, for a given constraint $s$ on the average distortion that can be introduced, the MU uses the optimal management policy to achieve the maximum minimal type II probability of error subject to a type I probability of error constraint $\varepsilon$, which, with slight abuse of notation, is denoted by $\beta(n,\varepsilon,s)$ as
\begin{equation}
\beta(n,\varepsilon,s)\triangleq\max_{\gamma^{n}(s)}\{\beta(n,\varepsilon,\gamma^{n}(s))\}.
\label{equation3}
\end{equation}

\begin{remark}
Different from the game theoretic formulation in \cite{barni2013,tondi2015,barni2016}, we take a worst-case analysis approach by assuming an informed AD who always performs the optimal hypothesis test. This is more appropriate for the privacy problem studied here and will lead to a privacy guarantee independent of the knowledge of the AD about the system.
\end{remark}

In the following, the optimal privacy-preserving policy is characterized in the asymptotic regime as $n\to\infty$, by focusing on the asymptotic exponent of the maximum minimal type II probability of error subject to a type I probability of error constraint.

We define the Kullback-Leibler divergence rate $\theta(s)$ as
\begin{equation}
\theta(s)\triangleq\inf_{k,\gamma^{k}(s)}\left\{\frac{1}{k}\textnormal{D}(p_{Y^{k}|h_{0}}||p_{Y^{k}|h_{1}})\right\},
\label{equation4}
\end{equation}
where the infimum is taken over all $k\in\mathbb{Z}_{+}$, and for each $k$, over all management policies that satisfy the average distortion constraint over a $k$-slot time horizon.

\begin{lemma}
\begin{equation*}
\theta(s)=\lim_{k\to\infty}\inf_{\gamma^{k}(s)}\left\{\frac{1}{k}\textnormal{D}(p_{Y^{k}|h_{0}}||p_{Y^{k}|h_{1}})\right\}.
\end{equation*}
\label{lemma1}
\end{lemma}

\begin{IEEEproof}
We first show that the following sequence $\left\{\inf_{\gamma^{k}(s)}\left\{\textnormal{D}(p_{Y^{k}|h_{0}}||p_{Y^{k}|h_{1}})\right\}\right\}_{k\in\mathbb{Z}_{+}}$ is {\it subadditive}. Given any $n$, $l\in\mathbb{Z}_{+}$, let $(\gamma^{n}(s),\gamma^{l}(s))$ denote a management policy, which uses $\gamma^{n}(s)$ over the first $n$ slots, and $\gamma^{l}(s)$ over the remaining $l$ slots. Therefore, $(\gamma^{n}(s),\gamma^{l}(s))$ satisfies the average distortion constraint over an $(n+l)$-slot time horizon. We have
\begin{equation*}
\begin{aligned}
&\inf_{\gamma^{n+l}(s)}\left\{\textnormal{D}(p_{Y^{n+l}|h_{0}}||p_{Y^{n+l}|h_{1}})\right\}\\
\leq&\inf_{(\gamma^{n}(s),\gamma^{l}(s))}\left\{\textnormal{D}(p_{Y^{n+l}|h_{0}}||p_{Y^{n+l}|h_{1}})\right\}\\
=&\inf_{\gamma^{n}(s)}\left\{\textnormal{D}(p_{Y^{n}|h_{0}}||p_{Y^{n}|h_{1}})\right\}+\inf_{\gamma^{l}(s)}\left\{\textnormal{D}(p_{Y^{l}|h_{0}}||p_{Y^{l}|h_{1}})\right\},
\end{aligned}
\end{equation*}
where the equality follows from the chain rule for the Kullback-Leibler divergence. Then, it follows from Fekete's Lemma \cite[Lemma 11.2]{csiszar2011} that
\begin{equation*}
\begin{aligned}
\theta(s)=&\inf_{k,\gamma^{k}(s)}\left\{\frac{1}{k}\textnormal{D}(p_{Y^{k}|h_{0}}||p_{Y^{k}|h_{1}})\right\}\\
=&\lim_{k\to\infty}\inf_{\gamma^{k}(s)}\left\{\frac{1}{k}\textnormal{D}(p_{Y^{k}|h_{0}}||p_{Y^{k}|h_{1}})\right\}.
\end{aligned}
\end{equation*}
\end{IEEEproof}

The following theorem characterizes the operational meaning of the Kullback-Leibler divergence rate $\theta(s)$ in the asymptotic assessment of the optimal privacy performance assuming an adversarial Neyman-Pearson hypothesis test.

\begin{theorem}
Given $s>0$,
\begin{equation}
\limsup_{n\to\infty}\frac{1}{n}\log\frac{1}{\beta(n,\varepsilon,s)}\leq\theta(s),\textnormal{ }\forall\varepsilon\in(0,1),
\label{equation5}
\end{equation}
and
\begin{equation}
\lim_{\varepsilon\to1}\liminf_{n\to\infty}\frac{1}{n}\log\frac{1}{\beta(n,\varepsilon,s)}\geq\theta(s).
\label{equation6}
\end{equation}
\label{theorem1}
\end{theorem}

\begin{IEEEproof}
Given any $k\in\mathbb{Z}_{+}$, $\gamma^{k}(s)$, and the resulting $p_{Y^{k}|h_{0}}$, $p_{Y^{k}|h_{1}}$, let $\gamma^{kl}(s)$ denote a management policy which repeatedly uses $\gamma^{k}(s)$ for $l$ times. From the definition in (\ref{equation3}) and Stein's Lemma \cite[Theorem 11.8.3]{cover2006}, it follows that
\begin{equation*}
\begin{aligned}
\limsup_{l\to\infty}\frac{1}{kl}\log\frac{1}{\beta(kl,\varepsilon,s)}\leq&\lim_{l\to\infty}\frac{1}{kl}\log\frac{1}{\beta(kl,\varepsilon,\gamma^{kl}(s))}\\
=&\frac{1}{k}\textnormal{D}(p_{Y^{k}|h_{0}}||p_{Y^{k}|h_{1}}),
\end{aligned}
\end{equation*}
for all $\varepsilon\in(0,1)$. For $k(l-1)<n\leq kl$, we have
\begin{equation*}
\beta(kl,\varepsilon,s)\leq\beta(n,\varepsilon,s)\leq\beta(k(l-1),\varepsilon,s).
\end{equation*}
It follows that
\begin{equation*}
\begin{aligned}
\limsup_{n\to\infty}\frac{1}{n}\log\frac{1}{\beta(n,\varepsilon,s)}&\leq\limsup_{l\to\infty}\frac{kl}{k(l-1)}\frac{1}{kl}\log\frac{1}{\beta(kl,\varepsilon,s)}\\
&=\limsup_{l\to\infty}\frac{1}{kl}\log\frac{1}{\beta(kl,\varepsilon,s)}\\
&\leq\frac{1}{k}\textnormal{D}(p_{Y^{k}|h_{0}}||p_{Y^{k}|h_{1}}),
\end{aligned}
\end{equation*}
for all $\varepsilon\in(0,1)$, $k\in\mathbb{Z}_{+}$, and $\gamma^{k}(s)$. Therefore, we have the upper bound
\begin{equation*}
\limsup_{n\to\infty}\frac{1}{n}\log\frac{1}{\beta(n,\varepsilon,s)}\leq\theta(s),\textnormal{ }\forall\varepsilon\in(0,1).
\end{equation*}

Given $0<\delta'<\infty$ and for all $n\in\mathbb{Z}_{+}$, let $\varepsilon'(n)\triangleq$
\begin{equation*}
\sup_{\gamma^{n}(s)}p_{Y^{n}|h_{0}}\left\{y^{n}\left|\log\frac{p_{Y^{n}|h_{0}}(y^{n})}{p_{Y^{n}|h_{1}}(y^{n})}<\textnormal{D}(p_{Y^{n}|h_{0}}||p_{Y^{n}|h_{1}})-\delta'\right.\right\},
\end{equation*}
and let $\epsilon(n)\triangleq$
\begin{equation*}
\sup_{\gamma^{n}(s)}p_{Y^{n}|h_{0}}\left\{y^{n}\left|\log\frac{p_{Y^{n}|h_{0}}(y^{n})}{p_{Y^{n}|h_{1}}(y^{n})}\leq\textnormal{D}(p_{Y^{n}|h_{0}}||p_{Y^{n}|h_{1}})-\delta'\right.\right\}.
\end{equation*}
Note that $\varepsilon'(n)\leq\epsilon(n)$. We further have $\epsilon(n)<1$ since $\epsilon(n)=1$ will lead to the contradiction $\textnormal{D}(p_{Y^{n}|h_{0}}||p_{Y^{n}|h_{1}})\leq\textnormal{D}(p_{Y^{n}|h_{0}}||p_{Y^{n}|h_{1}})-\delta'$.
It follows that $\varepsilon'(n)<1$. Suppose that $\gamma^{n*}(s)$ leads to $p_{Y^{n}|h_{0}}^{*}$, $p_{Y^{n}|h_{1}}^{*}$, and achieves $\beta(n,\varepsilon'(n),s)$. If the AD uses the following hypothesis testing strategy
\begin{equation}
\mathcal{A}_{n}=\left\{y^{n}\left|\frac{1}{n}\log\frac{p_{Y^{n}|h_{0}}^{*}(y^{n})}{p_{Y^{n}|h_{1}}^{*}(y^{n})}\geq t(n)\right.\right\},
\label{equation5.7}
\end{equation}
with the test threshold
\begin{equation}
t(n)=\frac{1}{n}\textnormal{D}(p_{Y^{n}|h_{0}}^{*}||p_{Y^{n}|h_{1}}^{*})-\frac{\delta'}{n},
\label{equation5.8}
\end{equation}
from the definition of $\varepsilon'(n)$, the corresponding type I probability of error satisfies the upper bound constraint
\begin{equation*}
p_{Y^{n}|h_{0}}^{*}(\mathcal{A}_{n}^{c})\leq\varepsilon'(n).
\end{equation*}
Since the hypothesis testing strategy in (\ref{equation5.7}) is not necessarily optimal for the AD, the definition of the maximum minimal type II probability of error implies that
\begin{equation}
\beta(n,\varepsilon'(n),s)\leq p_{Y^{n}|h_{1}}^{*}(\mathcal{A}_{n}).
\label{equation5.9}
\end{equation}
In \cite[Lemma 4.1.1]{han2003}, it has been shown that
\begin{equation}
p_{Y^{n}|h_{1}}^{*}(\mathcal{A}_{n})\leq\exp(-nt(n)).
\label{equation5.10}
\end{equation}
The inequalities (\ref{equation5.9}) and (\ref{equation5.10}) jointly lead to
\begin{equation*}
\begin{aligned}
&\beta(n,\varepsilon'(n),s)
\leq\exp(-nt(n))\\
\leq&\exp\left(-n\left(\inf_{\gamma^{n}(s)}\left\{\frac{1}{n}\textnormal{D}(p_{Y^{n}|h_{0}}||p_{Y^{n}|h_{1}})\right\}-\frac{\delta'}{n}\right)\right),
\end{aligned}
\end{equation*}
i.e., for all $n\in\mathbb{Z}_{+}$,
\begin{equation*}
\frac{1}{n}\log\frac{1}{\beta(n,\varepsilon'(n),s)}\geq\inf_{\gamma^{n}(s)}\left\{\frac{1}{n}\textnormal{D}(p_{Y^{n}|h_{0}}||p_{Y^{n}|h_{1}})\right\}-\frac{\delta'}{n}.
\end{equation*}
In the asymptotic regime as $n\to\infty$, it follows that
\begin{equation*}
\begin{aligned}
&\liminf_{n\to\infty}\frac{1}{n}\log\frac{1}{\beta(n,\varepsilon'(n),s)}\\
\geq&\lim_{n\to\infty}\inf_{\gamma^{n}(s)}\left\{\frac{1}{n}\textnormal{D}(p_{Y^{n}|h_{0}}||p_{Y^{n}|h_{1}})\right\}-\lim_{n\to\infty}\frac{\delta'}{n}=\theta(s),
\end{aligned}
\end{equation*}
where the equality follows from Lemma \ref{lemma1}. Note that $\liminf_{n\to\infty}\frac{1}{n}\log\frac{1}{\beta(n,\varepsilon,s)}$ is a monotone non-decreasing function of $\varepsilon$. Then the lower bound holds:
\begin{equation}
\begin{aligned}
\lim_{\varepsilon\to1}\liminf_{n\to\infty}\frac{1}{n}\log\frac{1}{\beta(n,\varepsilon,s)}\geq&\liminf_{n\to\infty}\frac{1}{n}\log\frac{1}{\beta(n,\varepsilon'(n),s)}\\
\geq&\theta(s).
\end{aligned}
\label{equation5.11}
\end{equation}
\end{IEEEproof}

When $\varepsilon$ is close to one, the bounds on the asymptotic exponent of the maximum minimal type II probability of error are tight, which is made more concrete in the following corollary.

\begin{corollary}
Given $s>0$,
\begin{equation*}
\lim_{\varepsilon\to1}\lim_{n\to\infty}\frac{1}{n}\log\frac{1}{\beta(n,\varepsilon,s)}=\theta(s).
\end{equation*}
\label{corollary1}
\end{corollary}

\begin{remark}
Given $s>0$, the case $\varepsilon\to1$ represents the worst privacy leakage scenario assuming the adversarial Neyman-Pearson hypothesis test, i.e., $\theta(s)$ is a privacy guarantee.
\label{remark1}
\end{remark}

In the following, we characterize the asymptotic privacy performances of two particular management policies in the worst case scenario, i.e., $\varepsilon\to1$.

\subsection{Memoryless Hypothesis-Aware Policy}
A simple MU might have a limited processing capability, and at time slot $i$, applies a memoryless hypothesis-aware management policy $\pi_{i}$, which can be time-variant, to determine the random adversarial observation $Y_{i}$ based on the current data $x_{i}$ and the true hypothesis $h$ as $Y_{i}=\pi_{i}(x_{i},h)$. Let $\pi^{n}\triangleq\{\pi_{i}\}_{i=1}^{n}:\mathcal{X}^{n}\times\mathcal{H}\to\mathcal{Y}^{n}$ denote a memoryless hypothesis-aware management policy over an $n$-slot time horizon. If $\pi^{n}$ satisfies the average distortion constraint in (\ref{equation2}), it is denoted by $\pi^{n}(s)$. When the MU adopts the optimal memoryless hypothesis-aware policy, the maximum minimal type II probability of error subject to a type I probability of error upper bound $\varepsilon$ is denoted by
\begin{equation}
\beta_{\textnormal{L}}(n,\varepsilon,s)\triangleq\max_{\pi^{n}(s)}\{\beta(n,\varepsilon,\pi^{n}(s))\}.
\label{equation54}
\end{equation}

We similarly define $\theta_{\textnormal{L}}(s)$ as
\begin{equation}
\theta_{\textnormal{L}}(s)\triangleq\inf_{k,\pi^{k}(s)}\left\{\frac{1}{k}\textnormal{D}(p_{Y^{k}|h_{0}}||p_{Y^{k}|h_{1}})\right\}.
\label{equation55}
\end{equation}
Following similar proof steps as in Theorem \ref{theorem1}, we can specify the asymptotic exponent of the maximum minimal type II probability of error by the Kullback-Leibler divergence rate $\theta_{\textnormal{L}}(s)$ when the MU adopts the optimal memoryless hypothesis-aware policy.
\begin{corollary}
Given $s>0$,
\begin{equation}
\lim_{\varepsilon\to1}\lim_{n\to\infty}\frac{1}{n}\log\frac{1}{\beta_{\textnormal{L}}(n,\varepsilon,s)}=\theta_{\textnormal{L}}(s).
\label{equation51}
\end{equation}
\label{corollary2}
\end{corollary}

We next show that the asymptotic exponent of the maximum minimal type II probability of error can also be characterized by a single-letter Kullback-Leibler divergence. To this end, we first define the following single-letter Kullback-Leibler divergence function of single-slot average distortion constraints under both hypotheses.

Given $\bar{s}$, $\tilde{s}>0$, we define $\phi(\bar{s},\tilde{s})$ as
\begin{equation}
\phi(\bar{s},\tilde{s})\triangleq\min_{(p_{Y|X,h_{0}},p_{Y|X,h_{1}})\in\mathcal{P}(\bar{s},\tilde{s})}\left\{\textnormal{D}(p_{Y|h_{0}}||p_{Y|h_{1}})\right\},
\label{equation20}
\end{equation}
where the minimization is over the convex domain
\begin{equation*}
\mathcal{P}(\bar{s},\tilde{s})
\triangleq\left\{(p_{Y|X,h_{0}},p_{Y|X,h_{1}})\left|\begin{gathered}\textnormal{E}[d(X,Y)|h_{0}]\leq\bar{s}\\\textnormal{E}[d(X,Y)|h_{1}]\leq\tilde{s}\end{gathered}\right.\right\}.
\end{equation*}
In the definition of $\mathcal{P}(\bar{s},\tilde{s})$, $\textnormal{E}[d(X,Y)|h_{0}]\leq\bar{s}$ denotes the single-slot average distortion constraint under hypothesis $h_{0}$; and $\textnormal{E}[d(X,Y)|h_{1}]\leq\tilde{s}$ denotes the single-slot average distortion constraint under hypothesis $h_{1}$.

\begin{lemma}
$\phi(\bar{s},\tilde{s})$ is a non-increasing, continuous, and jointly convex function for $\bar{s}>0$ and $\tilde{s}>0$.
\label{lemma2}
\end{lemma}

\begin{IEEEproof}
The non-increasing property of $\phi(\bar{s},\tilde{s})$ is self-evident. On the two-dimensional convex open set of $\bar{s}>0$, $\tilde{s}>0$, its continuity will follow from the convexity \cite{null1972}. Therefore, we only prove its convexity here. Assume that $(p_{Y|X,h_{0}}^{(1)},p_{Y|X,h_{1}}^{(1)})$ leads to $\phi(\bar{s}_{1},\tilde{s}_{1})=\textnormal{D}(p_{Y|h_{0}}^{(1)}||p_{Y|h_{1}}^{(1)})$ and $(p_{Y|X,h_{0}}^{(2)},p_{Y|X,h_{1}}^{(2)})$ leads to $\phi(\bar{s}_{2},\tilde{s}_{2})=\textnormal{D}(p_{Y|h_{0}}^{(2)}||p_{Y|h_{1}}^{(2)})$. For all $0\leq\lambda\leq1$,
\begin{equation*}
\begin{aligned}
&\lambda\phi(\bar{s}_{1},\tilde{s}_{1})+(1-\lambda)\phi(\bar{s}_{2},\tilde{s}_{2})\\
=&\lambda\textnormal{D}(p_{Y|h_{0}}^{(1)}||p_{Y|h_{1}}^{(1)})+(1-\lambda)\textnormal{D}(p_{Y|h_{0}}^{(2)}||p_{Y|h_{1}}^{(2)})\\
\geq&\textnormal{D}(\lambda p_{Y|h_{0}}^{(1)}+(1-\lambda)p_{Y|h_{0}}^{(2)}||\lambda p_{Y|h_{1}}^{(1)}+(1-\lambda)p_{Y|h_{1}}^{(2)})\\
\geq&\phi(\lambda\bar{s}_{1}+(1-\lambda)\bar{s}_{2},\lambda\tilde{s}_{1}+(1-\lambda)\tilde{s}_{2}),
\end{aligned}
\end{equation*}
where the first inequality follows from the convexity of $\textnormal{D}(\cdot||\cdot)$; and the second follows from the definition of $\phi(\bar{s},\tilde{s})$ in (\ref{equation20}).
\end{IEEEproof}

\begin{theorem}
Given $s>0$,
\begin{equation}
\theta_{\textnormal{L}}(s)=\phi(s,s).
\label{equation52}
\end{equation}
\label{theorem4}
\end{theorem}

\begin{IEEEproof}
For any $k\in\mathbb{Z}_{+}$, $\pi^{k}(s)$, and the resulting $p_{Y^{k}|h_{0}}$, $p_{Y^{k}|h_{1}}$, we have
\begin{equation*}
\begin{aligned}
&\frac{1}{k}\textnormal{D}\left(p_{Y^{k}|h_{0}}||p_{Y^{k}|h_{1}}\right)\\
\overset{(a)}{=}&\frac{1}{k}\sum_{i=1}^{k}\textnormal{D}\left(p_{Y_{i}|h_{0}}||p_{Y_{i}|h_{1}}\right)\\
\overset{(b)}{\geq}&\frac{1}{k}\sum_{i=1}^{k}\phi\left(\textnormal{E}[d(X_{i},Y_{i})|h_{0}],\textnormal{E}[d(X_{i},Y_{i})|h_{1}]\right)\\
\overset{(c)}{\geq}&\phi\left(\textnormal{E}\left[\frac{1}{k}\left.\sum_{i=1}^{k}d(X_{i},Y_{i})\right|h_{0}\right],\textnormal{E}\left[\frac{1}{k}\left.\sum_{i=1}^{k}d(X_{i},Y_{i})\right|h_{1}\right]\right)\\
\overset{(d)}{\geq}&\phi(s,s),
\end{aligned}
\end{equation*}
where $(a)$ follows since the policy $\pi^{k}(s)$ leads to $p_{Y^{k}|h_{j}}=\prod_{i=1}^{k}p_{Y_{i}|h_{j}}$ for $j=0,1$; $(b)$ follows from the definition of $\phi(\bar{s},\tilde{s})$; $(c)$ and $(d)$ follow from the convexity and the non-increasing property of $\phi(\bar{s},\tilde{s})$, respectively.

Therefore, we have
\begin{equation}
\theta_{\textnormal{L}}(s)=\inf_{k,\pi^{k}(s)}\left\{\frac{1}{k}\textnormal{D}(p_{Y^{k}|h_{0}}||p_{Y^{k}|h_{1}})\right\}\geq\phi(s,s).
\label{equation22}
\end{equation}

The proof of the opposite direction is straightforward. Let $(p_{Y|X,h_{0}}^{*},p_{Y|X,h_{1}}^{*})$ be the optimizer which achieves $\phi(s,s)$. It can be seen as a single-slot memoryless hypothesis-aware policy $\pi^{1}(s)$. From the definition of $\theta_{\textnormal{L}}(s)$ in (\ref{equation55}), it follows that
\begin{equation}
\theta_{\textnormal{L}}(s)\leq\phi(s,s).
\label{equation56}
\end{equation}
Alternatively, the inequality (\ref{equation56}) follows since $\phi(s,s)$ is the asymptotic exponent of the minimal type II probability of error achieved by a memoryless hypothesis-aware policy by using the single-slot policy $(p_{Y|X,h_{0}}^{*},p_{Y|X,h_{1}}^{*})$ at all slots.

The inequalities (\ref{equation22}) and (\ref{equation56}) jointly lead to Theorem \ref{theorem4}.
\end{IEEEproof}

\begin{remark}
Given $s>0$, the asymptotic exponent of the maximum minimal type II probability of error, $\lim_{\varepsilon\to1}\lim_{n\to\infty}\frac{1}{n}\log\frac{1}{\beta_{\textnormal{L}}(n,\varepsilon,s)}$, can be achieved by a memoryless hypothesis-aware policy which uses the single-slot policy $(p_{Y|X,h_{0}}^{*},p_{Y|X,h_{1}}^{*})$ corresponding to the optimizer of $\phi(s,s)$ at all time slots.
\end{remark}

\subsection{Hypothesis-Unaware Policy with Memory}
We now consider an MU which does not have access to the true hypothesis but has a large memory storage and a powerful processing capability. At time slot $i$, the MU follows a hypothesis-unaware management policy with memory, denoted by $\rho_{i}$, to determine the random adversarial observation $Y_{i}$ based on the data sequence $x^{i}$ and the past adversarial observations $y^{i-1}$ as $Y_{i}=\rho_{i}(x^{i},y^{i-1})$. Let $\rho^{n}\triangleq\{\rho_{i}\}_{i=1}^{n}:\mathcal{X}^{n}\to\mathcal{Y}^{n}$ denote a hypothesis-unaware management policy with memory over an $n$-slot time horizon. If $\rho^{n}$ satisfies the average distortion constraint in (\ref{equation2}), it is denoted by $\rho^{n}(s)$. When the MU uses the optimal privacy-preserving hypothesis-unaware policy with memory, the achieved maximum minimal type II probability of error subject to a type I probability of error upper bound $\varepsilon$ is denoted by
\begin{equation}
\beta_{\textnormal{M}}(n,\varepsilon,s)\triangleq\max_{\rho^{n}(s)}\{\beta(n,\varepsilon,\rho^{n}(s))\}.
\label{equation58}
\end{equation}

Similarly, we define a Kullback-Leibler divergence rate $\theta_{\textnormal{M}}(s)$ as
\begin{equation}
\theta_{\textnormal{M}}(s)\triangleq\inf_{k,\rho^{k}(s)}\left\{\frac{1}{k}\textnormal{D}(p_{Y^{k}|h_{0}}||p_{Y^{k}|h_{1}})\right\}.
\label{equation59}
\end{equation}
As specified in the following corollary, the asymptotic exponent of the maximum minimal type II probability of error can be characterized by the Kullback-Leibler divergence rate $\theta_{\textnormal{M}}(s)$ when the MU uses the optimal privacy-preserving hypothesis-unaware policy with memory.
\begin{corollary}
Given $s>0$,
\begin{equation}
\lim_{\varepsilon\to1}\lim_{n\to\infty}\frac{1}{n}\log\frac{1}{\beta_{\textnormal{M}}(n,\varepsilon,s)}=\theta_{\textnormal{M}}(s).
\label{equation60}
\end{equation}
\label{corollary3}
\end{corollary}

Compared with the memoryless hypothesis-aware policy, the hypothesis-unaware policy with memory has all the past data and adversarial observations while it does not know the true hypothesis. We next compare the asymptotic privacy performances of the two policies.
\begin{theorem}
Given $s>0$,
\begin{equation}
\theta_{\textnormal{M}}(s)\leq\theta_{\textnormal{L}}(s)=\phi(s,s).
\label{equation21}
\end{equation}
\label{theorem2}
\end{theorem}

To prove Theorem \ref{theorem2}, we will construct a two-phase hypothesis-unaware policy with memory, which first learns the hypothesis, and then we bound its privacy performance by $\theta_{\textnormal{M}}(s)$ and $\phi(s,s)$. The complete proof is given in the appendix.

\begin{remark}
The optimal privacy-preserving memoryless hypothesis-aware policy cannot outperform the optimal privacy-preserving hypothesis-unaware policy with memory. This is because the MU can learn the hypothesis with an arbitrarily small probability of error after observing a sufficiently long sequence of the original data.
\label{remark2}
\end{remark}

\section{Privacy-Preserving Data Management against a Bayesian Hypothesis Testing Adversary}
\label{section4}
In this section, we consider an adversarial Bayesian hypothesis testing formulation for the privacy problem. A particular Bayesian risk used here is the error probability of the AD. Thus, the minimal error probability of the AD measures the privacy leakage. The asymptotic privacy performance is studied under the adversarial Bayesian hypothesis testing framework.

The informed AD is assumed to use the optimal hypothesis testing strategy to achieve the minimal error probability as
\begin{equation*}
\alpha(n,\gamma^{n}(s))\triangleq\min_{\mathcal{A}_{n}\subseteq\mathcal{Y}^{n}}\left\{p_{0}\cdot p_{Y^{n}|h_{0}}(\mathcal{A}_{n}^{c})+p_{1}\cdot p_{Y^{n}|h_{1}}(\mathcal{A}_{n})\right\},
\end{equation*}
where $p_{0}$, $p_{1}$ denote the prior probabilities of hypotheses $h_{0}$, $h_{1}$; and $\mathcal{A}_{n}$, $\mathcal{A}_{n}^{c}$ denote the decision regions for $h_{0}$, $h_{1}$ of the AD. Correspondingly, the MU uses the optimal management policy that maximizes the minimal error probability of the AD as
\begin{equation}
\alpha(n,s)\triangleq\max_{\gamma^{n}(s)}\left\{\alpha(n,\gamma^{n}(s))\right\}.
\label{equation70}
\end{equation}

We will characterize the optimal privacy performance in the asymptotic regime as $n\to\infty$ by focusing on the asymptotic exponent of the maximum minimal error probability. To this end, we first introduce the Chernoff information. The Chernoff information of a probability distribution $P(Z)$ from another distribution $Q(Z)$ is defined as
\begin{equation*}
\textnormal{C}(P(Z),Q(Z))\triangleq\max_{0\leq\tau\leq1}\left\{\textnormal{C}_{\tau}(P(Z),Q(Z))\right\},
\end{equation*}
where
\begin{equation*}
\textnormal{C}_{\tau}(P(Z),Q(Z))\triangleq-\log\left(\sum_{z\in\mathcal{Z}}P^{\tau}(z)Q^{1-\tau}(z)\right).
\end{equation*}
The convexity of Chernoff information is shown in the following propositions.

\begin{proposition}
Given $0\leq\tau\leq1$, the function $\textnormal{C}_{\tau}(P(Z),Q(Z))$ is jointly convex in $P(Z)$ and $Q(Z)$.
\label{proposition1}
\end{proposition}

\begin{IEEEproof}
Given $0\leq\tau<1$, the function $\textnormal{C}_{\tau}(P(Z),Q(Z))$ is related to the R\'{e}nyi divergence $\textnormal{D}_{\tau}(P(Z)||Q(Z))$ as
\begin{equation*}
\textnormal{C}_{\tau}(P(Z),Q(Z))=(1-\tau)\cdot\textnormal{D}_{\tau}(P(Z)||Q(Z)).
\end{equation*}
The convexity of $\textnormal{C}_{\tau}(P(Z),Q(Z))$ follows from the convexity of R\'{e}nyi divergence \cite[Theorem 11]{tim2014} since $1-\tau$ is a positive scalar.

If $\tau=1$, we have
\begin{equation*}
\begin{aligned}
\textnormal{C}_{1}(P(Z),Q(Z))=&-\log\left(P\left(\left\{z\in\mathcal{Z}|Q(z)>0\right\}\right)\right)\\
=&\textnormal{D}_{0}(Q(Z)||P(Z)).
\end{aligned}
\end{equation*}
In this case, the convexity of $\textnormal{C}_{1}(P(Z),Q(Z))$ follows from the convexity of $\textnormal{D}_{0}(Q(Z)||P(Z))$.
\end{IEEEproof}

\begin{proposition}
The Chernoff information $\textnormal{C}(P(Z),Q(Z))$ is jointly convex in $P(Z)$ and $Q(Z)$.
\label{proposition2}
\end{proposition}

The convexity of $\textnormal{C}(P(Z),Q(Z))$ follows from the convexity of $\textnormal{C}_{\tau}(P(Z),Q(Z))$ for all $0\leq\tau\leq1$ and the fact that pointwise maximum preserves convexity \cite[Section 3.2.3]{boyd2004}.

We define the Chernoff information rate $\mu(s)$ as follows:
\begin{equation}
\mu(s)\triangleq\inf_{k,\gamma^{k}(s)}\left\{\frac{1}{k}\textnormal{C}(p_{Y^{k}|h_{0}},p_{Y^{k}|h_{1}})\right\}.
\label{equation71}
\end{equation}
The following lemma shows that the infimum over $k\in\mathbb{Z}_{+}$ in the definition of the Chernoff information rate $\mu(s)$ is taken at the limit.

\begin{lemma}
\begin{equation*}
\mu(s)=\lim_{k\to\infty}\inf_{\gamma^{k}(s)}\left\{\frac{1}{k}\textnormal{C}(p_{Y^{k}|h_{0}},p_{Y^{k}|h_{1}})\right\}.
\end{equation*}
\label{lemma5}
\end{lemma}

\begin{IEEEproof}
Given any $n$, $l\in\mathbb{Z}_{+}$, as defined before, $(\gamma^{n}(s),\gamma^{l}(s))$ is a management policy satisfying the average distortion constraint over an $(n+l)$-slot time horizon. We have
\begin{equation*}
\begin{aligned}
&\inf_{\gamma^{n+l}(s)}\left\{\textnormal{C}(p_{Y^{n+l}|h_{0}},p_{Y^{n+l}|h_{1}})\right\}\\
\leq&\inf_{(\gamma^{n}(s),\gamma^{l}(s))}\left\{\textnormal{C}(p_{Y^{n+l}|h_{0}},p_{Y^{n+l}|h_{1}})\right\}\\
\overset{(a)}{=}&\inf_{(\gamma^{n}(s),\gamma^{l}(s))}\{\max_{0\leq\tau\leq1}\{\textnormal{C}_{\tau}(p_{Y^{n}|h_{0}},p_{Y^{n}|h_{1}})\\
&+\textnormal{C}_{\tau}(p_{Y^{l}|h_{0}},p_{Y^{l}|h_{1}})\}\}\\
\leq&\inf_{(\gamma^{n}(s),\gamma^{l}(s))}\{\max_{0\leq\kappa\leq1}\{\textnormal{C}_{\kappa}(p_{Y^{n}|h_{0}},p_{Y^{n}|h_{1}})\}\\
&+\max_{0\leq\sigma\leq1}\{\textnormal{C}_{\sigma}(p_{Y^{l}|h_{0}},p_{Y^{l}|h_{1}})\}\}\\
=&\inf_{\gamma^{n}(s)}\left\{\textnormal{C}(p_{Y^{n}|h_{0}},p_{Y^{n}|h_{1}})\right\}+\inf_{\gamma^{l}(s)}\left\{\textnormal{C}(p_{Y^{l}|h_{0}},p_{Y^{l}|h_{1}})\right\},
\end{aligned}
\end{equation*}
where $(a)$ follows from the independence property $p_{Y^{n+l}|h_{j}}=p_{Y^{n}|h_{j}}\cdot p_{Y^{l}|h_{j}}$, $j=0,1$, satisfied by the management policy $(\gamma^{n}(s),\gamma^{l}(s))$. Therefore, the sequence $\left\{\inf_{\gamma^{k}(s)}\left\{\textnormal{C}(p_{Y^{k}|h_{0}},p_{Y^{k}|h_{1}})\right\}\right\}_{k\in\mathbb{Z}_{+}}$ is {\it subadditive}. Then, Lemma \ref{lemma5} follows from Fekete's Lemma.
\end{IEEEproof}

Next, we show that the asymptotic exponent of the maximum minimal error probability is characterized by the Chernoff information rate $\mu(s)$.

\begin{theorem}
Given $s>0$,
\begin{equation}
\lim_{n\to\infty}\frac{1}{n}\log\frac{1}{\alpha(n,s)}=\mu(s).
\label{equation72}
\end{equation}
\label{theorem5}
\end{theorem}

\begin{IEEEproof}
Given any $k\in\mathbb{Z}_{+}$, $\gamma^{k}(s)$, and the resulting $p_{Y^{k}|h_{0}}$, $p_{Y^{k}|h_{1}}$, as defined before, $\gamma^{kl}(s)$ is a management policy which repeatedly uses $\gamma^{k}(s)$ for $l$ times and satisfies the average distortion constraint over a $kl$-slot time horizon. From the optimality in the definition (\ref{equation70}) and the theorem \cite[Theorem 11.9.1]{cover2006}, it follows that
\begin{equation*}
\begin{aligned}
\limsup_{l\to\infty}\frac{1}{kl}\log\frac{1}{\alpha(kl,s)}\leq&\lim_{l\to\infty}\frac{1}{kl}\log\frac{1}{\alpha(kl,\gamma^{kl}(s))}\\
=&\frac{1}{k}\textnormal{C}(p_{Y^{k}|h_{0}},p_{Y^{k}|h_{1}}).
\end{aligned}
\end{equation*}
For $k(l-1)<n\leq kl$, we have
\begin{equation*}
\alpha(kl,s)\leq\alpha(n,s)\leq\alpha(k(l-1),s).
\end{equation*}
It follows that
\begin{equation*}
\begin{aligned}
\limsup_{n\to\infty}\frac{1}{n}\log\frac{1}{\alpha(n,s)}&\leq\limsup_{l\to\infty}\frac{kl}{k(l-1)}\frac{1}{kl}\log\frac{1}{\alpha(kl,s)}\\
&=\limsup_{l\to\infty}\frac{1}{kl}\log\frac{1}{\alpha(kl,s)}\\
&\leq\frac{1}{k}\textnormal{C}(p_{Y^{k}|h_{0}},p_{Y^{k}|h_{1}}),
\end{aligned}
\end{equation*}
for all $k\in\mathbb{Z}_{+}$ and $\gamma^{k}(s)$. Therefore, we have the upper bound
\begin{equation}
\limsup_{n\to\infty}\frac{1}{n}\log\frac{1}{\alpha(n,s)}\leq\mu(s).
\label{equation73}
\end{equation}
Given any $n\in\mathbb{Z}_{+}$, suppose that $\gamma^{n*}(s)$ leads to $p_{Y^{n}|h_{0}}^{*}$, $p_{Y^{n}|h_{1}}^{*}$, and achieves $\alpha(n,s)$. An optimal hypothesis testing strategy of the AD is a deterministic likelihood-ratio test \cite{trees2013} given by
\begin{equation*}
\mathcal{A}_{n}^{*}=\left\{y^{n}\left|\frac{p_{Y^{n}|h_{0}}^{*}(y^{n})}{p_{Y^{n}|h_{1}}^{*}(y^{n})}\geq\frac{p_{1}}{p_{0}}\right.\right\}.
\end{equation*}
Based on the optimal strategy of the AD, we can then rewrite the maximum minimal error probability $\alpha(n,s)$, and derive upper bounds on it as follows. For all $0\leq\tau\leq1$,
\begin{equation*}
\begin{aligned}
\alpha(n,s)=&\sum_{\mathcal{Y}^{n}}\min\left\{p_{0}\cdot p_{Y^{n}|h_{0}}^{*}(y^{n}),p_{1}\cdot p_{Y^{n}|h_{1}}^{*}(y^{n})\right\}\\
\overset{(a)}{\leq}&\sum_{\mathcal{Y}^{n}}p_{0}^{\tau}\cdot p_{Y^{n}|h_{0}}^{*\tau}(y^{n})\cdot p_{1}^{1-\tau}\cdot p_{Y^{n}|h_{1}}^{*1-\tau}(y^{n})\\
\leq&\sum_{\mathcal{Y}^{n}}p_{Y^{n}|h_{0}}^{*\tau}(y^{n})p_{Y^{n}|h_{1}}^{*1-\tau}(y^{n}),
\end{aligned}
\end{equation*}
where the inequality $(a)$ follows from \cite[(11.244)]{cover2006}. Therefore, we have
\begin{equation*}
\begin{aligned}
&\frac{1}{n}\log\frac{1}{\alpha(n,s)}\\
\geq&\frac{1}{n}\max_{0\leq\tau\leq1}\left\{-\log\left(\sum_{\mathcal{Y}^{n}}p_{Y^{n}|h_{0}}^{*\tau}(y^{n})p_{Y^{n}|h_{1}}^{*1-\tau}(y^{n})\right)\right\}\\
=&\frac{1}{n}\textnormal{C}(p_{Y^{n}|h_{0}}^{*},p_{Y^{n}|h_{1}}^{*})\\
\geq&\inf_{\gamma^{n}(s)}\left\{\frac{1}{n}\textnormal{C}(p_{Y^{n}|h_{0}},p_{Y^{n}|h_{1}})\right\}.
\end{aligned}
\end{equation*}
In the asymptotic regime, as $n\to\infty$, we have
\begin{equation}
\begin{aligned}
\liminf_{n\to\infty}\frac{1}{n}\log\frac{1}{\alpha(n,s)}\geq&\lim_{n\to\infty}\inf_{\gamma^{n}(s)}\left\{\frac{1}{n}\textnormal{C}(p_{Y^{n}|h_{0}},p_{Y^{n}|h_{1}})\right\}\\
=&\mu(s),
\end{aligned}
\label{equation74}
\end{equation}
where the final equality follows from Lemma \ref{lemma5}.

The inequalities (\ref{equation73}) and (\ref{equation74}) jointly lead to Theorem \ref{theorem5}.
\end{IEEEproof}

\begin{remark}
Over a finite time horizon, the prior distribution of the hypothesis determines the test threshold of the optimal likelihood-ratio test of the AD, and further determines the exponent of the maximum minimal error probability. However, as shown in Theorem \ref{theorem5}, the asymptotic exponent of the maximum minimal error probability does not depend on the prior distribution.
\label{remark5}
\end{remark}

In the following, we characterize the privacy-preserving managements for the two particular cases, the memoryless hypothesis-aware policy and the hypothesis-unaware policy with memory, under the adversarial Bayesian hypothesis testing framework.

\subsection{Memoryless Hypothesis-Aware Policy}
When the MU uses the optimal memoryless hypothesis-aware policy $\pi^{n*}(s)$ against the adversarial Bayesian hypothesis test, the achieved maximum minimal error probability is denoted by
\begin{equation}
\alpha_{\textnormal{L}}(n,s)\triangleq\max_{\pi^{n}(s)}\{\alpha(n,\pi^{n}(s))\}.
\label{equation75}
\end{equation}

We similarly define $\mu_{\textnormal{L}}(s)$ as
\begin{equation}
\mu_{\textnormal{L}}(s)\triangleq\inf_{k,\pi^{k}(s)}\left\{\frac{1}{k}\textnormal{C}(p_{Y^{k}|h_{0}},p_{Y^{k}|h_{1}})\right\}.
\label{equation76}
\end{equation}
Following similar proof steps as in Theorem \ref{theorem5}, we can show that the asymptotic exponent of the maximum minimal error probability is specified by the Chernoff information rate $\mu_{\textnormal{L}}(s)$ when the MU uses the optimal privacy-preserving memoryless hypothesis-aware policy.
\begin{corollary}
Given $s>0$,
\begin{equation}
\lim_{n\to\infty}\frac{1}{n}\log\frac{1}{\alpha_{\textnormal{L}}(n,s)}=\mu_{\textnormal{L}}(s).
\label{equation77}
\end{equation}
\label{corollary5}
\end{corollary}

We next define two single-letter expressions, which characterize upper and lower bounds on the asymptotic exponent of the maximum minimal error probability. Given $\bar{s}$, $\tilde{s}>0$ and $0\leq\tau\leq1$, we define
\begin{equation}
\nu_{\tau}(\bar{s},\tilde{s})\triangleq\min_{(p_{Y|X,h_{0}},p_{Y|X,h_{1}})\in\mathcal{P}(\bar{s},\tilde{s})}\left\{\textnormal{C}_{\tau}(p_{Y|h_{0}},p_{Y|h_{1}})\right\},
\label{equation78}
\end{equation}
and define the single-letter Chernoff information as
\begin{equation}
\nu(\bar{s},\tilde{s})\triangleq\min_{(p_{Y|X,h_{0}},p_{Y|X,h_{1}})\in\mathcal{P}(\bar{s},\tilde{s})}\left\{\textnormal{C}(p_{Y|h_{0}},p_{Y|h_{1}})\right\}.
\label{equation80}
\end{equation}

\begin{lemma}
Given $0\leq\tau\leq1$, $\nu_{\tau}(\bar{s},\tilde{s})$ is a non-increasing and jointly convex function for $\bar{s}>0$ and $\tilde{s}>0$.
\label{lemma6}
\end{lemma}

\begin{lemma}
$\nu(\bar{s},\tilde{s})$ is a non-increasing, continuous, and jointly convex function for $\bar{s}>0$ and $\tilde{s}>0$.
\label{lemma8}
\end{lemma}

The proofs of Lemmas \ref{lemma6} and \ref{lemma8} follow from the same arguments as in the proof of Lemma \ref{lemma2}, and are therefore omitted.

\begin{lemma}
Given $s>0$, we have
\begin{equation}
\max_{0\leq\tau\leq1}\left\{\nu_{\tau}(s,s)\right\}\leq\mu_{\textnormal{L}}(s)\leq\nu(s,s).
\label{equation79}
\end{equation}
\label{lemma7}
\end{lemma}

\begin{IEEEproof}
Given any $0\leq\tau\leq1$, $k\in\mathbb{Z}_{+}$, memoryless hypothesis-aware policy $\pi^{k}(s)$, and the resulting $p_{Y^{k}|h_{0}}$, $p_{Y^{k}|h_{1}}$, we have
\begin{equation}
\begin{aligned}
&\frac{1}{k}\textnormal{C}(p_{Y^{k}|h_{0}},p_{Y^{k}|h_{1}})\\
=&\frac{1}{k}\max_{0\leq\kappa\leq1}\left\{\textnormal{C}_{\kappa}\left(p_{Y^{k}|h_{0}},p_{Y^{k}|h_{1}}\right)\right\}\\
\geq&\frac{1}{k}\textnormal{C}_{\tau}\left(p_{Y^{k}|h_{0}},p_{Y^{k}|h_{1}}\right)\\
\overset{(a)}{=}&\frac{1}{k}\sum_{i=1}^{k}\textnormal{C}_{\tau}\left(p_{Y_{i}|h_{0}},p_{Y_{i}|h_{1}}\right)\\
\overset{(b)}{\geq}&\frac{1}{k}\sum_{i=1}^{k}\nu_{\tau}\left(\textnormal{E}[d(X_{i},Y_{i})|h_{0}],\textnormal{E}[d(X_{i},Y_{i})|h_{1}]\right)\\
\overset{(c)}{\geq}&\nu_{\tau}\left(\textnormal{E}\left[\frac{1}{k}\left.\sum_{i=1}^{k}d(X_{i},Y_{i})\right|h_{0}\right],\textnormal{E}\left[\frac{1}{k}\left.\sum_{i=1}^{k}d(X_{i},Y_{i})\right|h_{1}\right]\right)\\
\overset{(d)}{\geq}&\nu_{\tau}(s,s),
\end{aligned}
\label{equation81}
\end{equation}
where $(a)$ follows since the policy $\pi^{k}(s)$ leads to $p_{Y^{k}|h_{j}}=\prod_{i=1}^{k}p_{Y_{i}|h_{j}}$ for $j=0,1$; $(b)$ follows from the definition of $\nu_{\tau}(\bar{s},\tilde{s})$; $(c)$ follows from the convexity of $\nu_{\tau}(\bar{s},\tilde{s})$; and finally $(d)$ follows from the non-increasing property of $\nu_{\tau}(\bar{s},\tilde{s})$.

For any $k\in\mathbb{Z}_{+}$, $\pi^{k}(s)$, and the resulting $p_{Y^{k}|h_{0}}$, $p_{Y^{k}|h_{1}}$, we have
\begin{equation*}
\frac{1}{k}\textnormal{C}(p_{Y^{k}|h_{0}},p_{Y^{k}|h_{1}})\geq\max_{0\leq\tau\leq1}\left\{\nu_{\tau}(s,s)\right\},
\end{equation*}
since the inequality (\ref{equation81}) holds for all $0\leq\tau\leq1$. It further follows that
\begin{equation*}
\mu_{\textnormal{L}}(s)=\inf_{k,\pi^{k}(s)}\left\{\frac{1}{k}\textnormal{C}(p_{Y^{k}|h_{0}},p_{Y^{k}|h_{1}})\right\}\geq\max_{0\leq\tau\leq1}\left\{\nu_{\tau}(s,s)\right\}.
\end{equation*}

The other inequality $\mu_{\textnormal{L}}(s)\leq\nu(s,s)$ in (\ref{equation79}) follows from the definitions of $\mu_{\textnormal{L}}(s)$ and $\nu(s,s)$.
\end{IEEEproof}

Under the adversarial Bayesian hypothesis test setting, we have obtained a $\max\min$ single-letter lower bound and a $\min\max$ single-letter upper bound on the asymptotic exponent of the maximum minimal error probability when we focus on memoryless hypothesis-aware policy. In the following theorem, we show that the two bounds match each other, and the asymptotic exponent of the maximum minimal error probability can be specified by the single-letter Chernoff information $\nu(s,s)$.

\begin{theorem}
Given $s>0$,
\begin{equation}
\mu_{\textnormal{L}}(s)=\nu(s,s).
\label{equation82}
\end{equation}
\label{theorem6}
\end{theorem}

\begin{IEEEproof}
Given $s>0$, the lower and upper bounds derived in Lemma \ref{lemma7} can be specified by a $\max\min$ expression and a $\min\max$ expression as follows:
\begin{equation*}
\begin{aligned}
&\max_{0\leq\tau\leq1}\{\nu_{\tau}(s,s)\}\\
=&\max_{0\leq\tau\leq1}\left\{\min_{(p_{Y|X,h_{0}},p_{Y|X,h_{1}})\in\mathcal{P}(s,s)}\left\{\textnormal{C}_{\tau}(p_{Y|h_{0}},p_{Y|h_{1}})\right\}\right\},
\end{aligned}
\end{equation*}
\begin{equation*}
\begin{aligned}
&\nu(s,s)\\
=&\min_{(p_{Y|X,h_{0}},p_{Y|X,h_{1}})\in\mathcal{P}(s,s)}\left\{\max_{0\leq\tau\leq1}\left\{\textnormal{C}_{\tau}(p_{Y|h_{0}},p_{Y|h_{1}})\right\}\right\}.
\end{aligned}
\end{equation*}
If $\tau$ is fixed, $\textnormal{C}_{\tau}(p_{Y|h_{0}},p_{Y|h_{1}})$ is a jointly convex function in $p_{Y|X,h_{0}}$ and $p_{Y|X,h_{1}}$, which follows from the convexity of $\textnormal{C}_{\tau}(\cdot,\cdot)$ shown in Proposition \ref{proposition1} and the convexity-preserving composition rule in \cite[Section 3.2.4]{boyd2004}. If $p_{Y|X,h_{0}}$ and $p_{Y|X,h_{1}}$ are fixed, $p_{Y|h_{0}}$ and $p_{Y|h_{1}}$ are fixed, and $\textnormal{C}_{\tau}(p_{Y|h_{0}},p_{Y|h_{1}})$ is a concave function in $\tau$, which follows from the result \cite[Corollary 2]{tim2014}.

From von Neumann's Minimax Theorem \cite{hukukane1954}, it follows that
\begin{equation}
\max_{0\leq\tau\leq1}\{\nu_{\tau}(s,s)\}=\nu(s,s).
\label{equation83}
\end{equation}
Lemma \ref{lemma7} and (\ref{equation83}) jointly lead to Theorem \ref{theorem6}.
\end{IEEEproof}

\begin{remark}
Given $s>0$, the asymptotic exponent of the maximum minimal error probability, $\lim_{n\to\infty}\frac{1}{n}\log\frac{1}{\alpha_{\textnormal{L}}(n,s)}$, can be achieved by a memoryless hypothesis-aware policy which uses the single-slot policy $(p_{Y|X,h_{0}}^{*},p_{Y|X,h_{1}}^{*})$ corresponding to the optimizer of $\nu(s,s)$ at all time slots.
\label{remark6}
\end{remark}

\subsection{Hypothesis-Unaware Policy with Memory}
When the MU uses the optimal hypothesis-unaware policy with memory, the achieved maximum minimal error probability is denoted by
\begin{equation}
\alpha_{\textnormal{M}}(n,s)\triangleq\max_{\rho^{n}(s)}\left\{\alpha(n,\rho^{n}(s))\right\}.
\label{equation84}
\end{equation}

We define the Chernoff information rate $\mu_{\textnormal{M}}(s)$ as
\begin{equation}
\mu_{\textnormal{M}}(s)\triangleq\inf_{k,\rho^{k}(s)}\left\{\frac{1}{k}\textnormal{C}\left(p_{Y^{k}|h_{0}},p_{Y^{k}|h_{1}}\right)\right\},
\label{equation85}
\end{equation}
which characterizes the asymptotic exponent of the maximum minimal error probability as shown in the following corollary.

\begin{corollary}
Given $s>0$,
\begin{equation}
\lim_{n\to\infty}\frac{1}{n}\log\frac{1}{\alpha_{\textnormal{M}}(n,s)}=\mu_{\textnormal{M}}(s).
\label{equation86}
\end{equation}
\label{corollary6}
\end{corollary}

Similar to the adversarial Neyman-Pearson hypothesis testing case, the following theorem shows that the optimal memoryless hypothesis-aware policy cannot outperform the optimal hypothesis-unaware policy with memory against the adversarial Bayesian hypothesis test.

\begin{theorem}
Given $s>0$,
\begin{equation}
\mu_{\textnormal{M}}(s)\leq\mu_{\textnormal{L}}(s)=\nu(s,s).
\label{equation87}
\end{equation}
\label{theorem7}
\end{theorem}

The proof of Theorem \ref{theorem7} follows similarly to Theorem \ref{theorem2}. We can construct a two-phase hypothesis-unaware policy with memory, which first learns the hypothesis. Then, we bound its privacy performance with $\mu_{\textnormal{M}}(s)$ and $\nu(s,s)$. The complete proof is given in the appendix.

\begin{figure}
\centering
\includegraphics[scale=0.45]{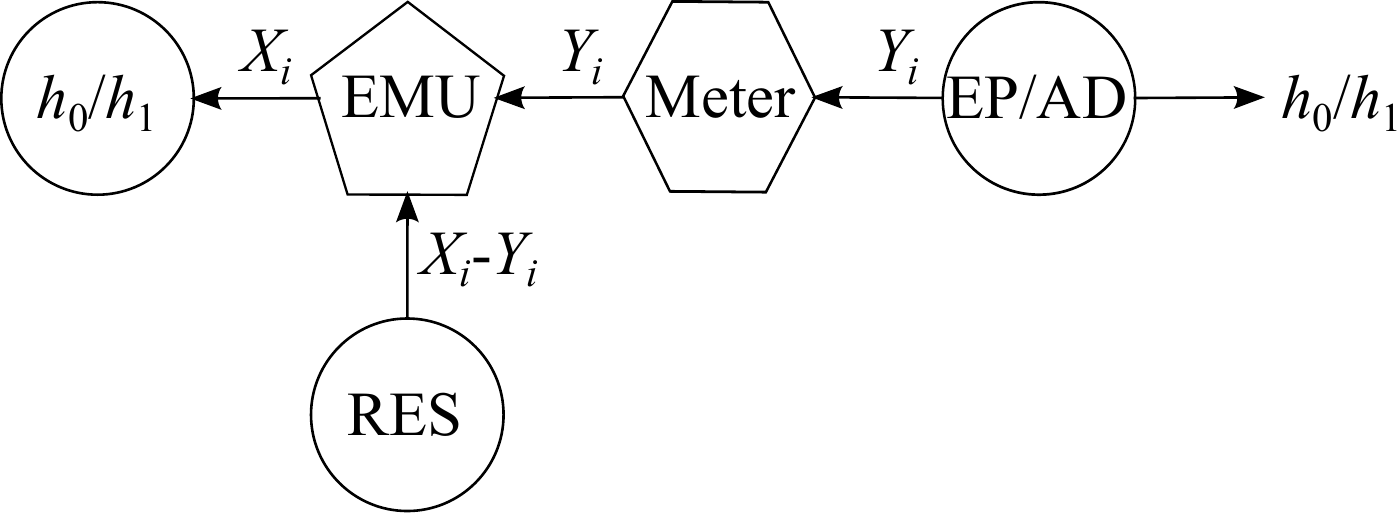}
\caption{The smart meter privacy problem in the presence of a renewable energy source (RES).}
\label{figure20}
\end{figure}

\section{Application to Smart Meter Privacy Problem}
\label{section5}
In this section, we consider an application of the theoretical framework we have introduced above to the smart meter privacy problem.

\subsection{Smart Meter Privacy Model}
The considered smart meter privacy problem is shown in Fig. \ref{figure20}. The i.i.d. data $X_{i}$ denotes the {\it non-negative} energy demand at time slot $i$, from a finite energy demand alphabet $\mathcal{X}$. The binary hypothesis represents the private user behavior, e.g., the user is at home or not, or a particular appliance is being used or not. Assuming that the EP is an AD, the adversarial observation $Y_{i}$ corresponds in this setting to the {\it non-negative} meter reading, or equivalently the energy supply from the EP, at time slot $i$. The finite energy supply alphabet satisfies $\mathcal{Y} \supseteq \mathcal{X}$.

The energy management unit (EMU) is a processor run by the user that manages the energy supplies to satisfy the energy demand. At every time slot, the EMU follows an instantaneous energy management policy to determine $Y_{i}$. It is assumed that the rest of the energy demand, $X_{i}-Y_{i}$, is satisfied from the renewable energy source (RES). We impose the following instantaneous constraint:
\begin{equation}
p_{Y_{i}|X_{i}}(y_{i}|x_{i})=0,\;\textnormal{if}\;y_{i}>x_{i},
\label{equation1}
\end{equation}
i.e., the RES cannot be charged with energy from the EP. We assume that the RES has a positive average energy generation rate $s$ and is equipped with a sufficiently large energy storage. Accordingly, the following average energy constraint is imposed on the energy management policy over an $n$-slot time horizon for the long-term availability of renewable energy supply:
\begin{equation}
\textnormal{E}\left[\left.\frac{1}{n}\sum_{i=1}^{n}(X_{i}-Y_{i})\right|h_{j}\right]\leq s,\textnormal{ }j=0,1.
\label{equation30}
\end{equation}

The assumptions on the informed AD and the adversarial hypothesis test are the same as the general problem. Here the EP is considered as an AD, who has the authority to access the meter readings of energy supplies from the EP, and has all the statistical knowledge, but does not have access to the amount of energy supply from the RES, which is available locally to the user.

The privacy-preserving energy management is to modify the energy demand sequence $\{X_i\}$ into an energy supply sequence $\{Y_i\}$ by exploiting the RES in an online manner to prevent the EP from correctly inferring the user behavior.

\subsection{Privacy-Preserving Energy Management}
Note that an energy management policy satisfying (\ref{equation1}) and (\ref{equation30}) will satisfy the constraint (\ref{equation2}) in the general problem. Therefore, the obtained theoretic results of asymptotic privacy performance in the general problem can be directly applied to the smart meter privacy problem. Here, we will focus on the memoryless hypothesis-aware energy management policy, which can be easily designed and implemented.

Under the adversarial Neyman-Pearson hypothesis test, and given a renewable energy generation rate $s>0$, it follows from Corollary \ref{corollary2} and Theorem \ref{theorem4} that the exponent of the maximum minimal type II probability of error can be characterized by
\begin{equation}
\phi(s,s)=\min_{(p_{Y|X,h_{0}},p_{Y|X,h_{1}})\in\mathcal{P_{\textnormal{E}}}(s,s)}\left\{\textnormal{D}(p_{Y|h_{0}}||p_{Y|h_{1}})\right\},
\label{equation31}
\end{equation}
with the convex set
\begin{equation*}
\begin{aligned}
&\mathcal{P}_{\textnormal{E}}(s,s)\\
\triangleq&\left\{(p_{Y|X,h_{0}},p_{Y|X,h_{1}})\left|\begin{gathered}\textnormal{E}[X-Y|h_{0}]\leq s\\\textnormal{E}[X-Y|h_{1}]\leq s\\p_{Y|X,h_{0}}(y|x)=0,\,\textnormal{if}\,y>x\\p_{Y|X,h_{1}}(y|x)=0,\,\textnormal{if}\,y>x\end{gathered}\right.\right\}.
\end{aligned}
\end{equation*}
While solving the optimization problem in (\ref{equation31}) leads to the optimal privacy performance, the energy supply alphabet $\mathcal{Y}$ can be very large, which means a highly complex optimization problem. On the other hand, the energy demand alphabet $\mathcal{X}$ is determined by a number of operation modes of the appliances and is typically finite. We show in the next theorem that the alphabet $\mathcal{Y}$ can be limited to the alphabet $\mathcal{X}$. This result can greatly simplify the numerical evaluation of $\phi(s,s)$ in the considered smart meter privacy context.

\begin{theorem}
The energy supply alphabet can be limited to the energy demand alphabet under both hypotheses without loss of optimality for the evaluation of $\phi(s,s)$.
\label{theorem3}
\end{theorem}

\begin{IEEEproof}
Suppose that $\phi(s,s)=\textnormal{D}(p_{Y|h_{0}}^{*}||p_{Y|h_{1}}^{*})$ is achieved by $p_{Y|X,h_{0}}^{*}$ and $p_{Y|X,h_{1}}^{*}$. Let $\mathcal{X}=\{x_{(1)},\dots,x_{(|\mathcal{X}|)}\}$ with $x_{(i)}<x_{(k)}$ if $i<k$. Consider the following quantization operation which maps $y$ to $\hat{y}$:
\begin{equation*}
\hat{y}=\left\{\begin{aligned}x_{(i)},&\textnormal{ if }y\in(x_{(i-1)},x_{(i)}],\textnormal{ }i\geq2\\x_{(1)},&\textnormal{ if }y\in[0,x_{(1)}]\end{aligned}\right..
\end{equation*}
It can be verified that $p_{\hat{Y}|X,h_{0}}$, $p_{\hat{Y}|X,h_{1}}$ satisfy the constraints in the definition of $\mathcal{P}_{\textnormal{E}}(s,s)$. From the optimality in the definition of $\phi(s,s)$, we have
\begin{equation*}
\phi(s,s)=\textnormal{D}(p_{Y|h_{0}}^{*}||p_{Y|h_{1}}^{*})\leq\textnormal{D}(p_{\hat{Y}|h_{0}}||p_{\hat{Y}|h_{1}}).
\end{equation*}
In addition, due to the data processing inequality of Kullback-Leibler divergence \cite[Theorem 9]{tim2014}, we have
\begin{equation*}
\phi(s,s)=\textnormal{D}(p_{Y|h_{0}}^{*}||p_{Y|h_{1}}^{*})\geq\textnormal{D}(p_{\hat{Y}|h_{0}}||p_{\hat{Y}|h_{1}}).
\end{equation*}
Therefore,
\begin{equation*}
\phi(s,s)=\textnormal{D}(p_{\hat{Y}|h_{0}}||p_{\hat{Y}|h_{1}}),
\end{equation*}
and the energy supply alphabet under both hypotheses can be constrained to $\mathcal{X}$ without loss of optimality.
\end{IEEEproof}

Under the adversarial Bayesian hypothesis test, and given a renewable energy generation rate $s >0$, it follows from Corollary \ref{corollary5} and Theorem \ref{theorem6} that the exponent of the maximum minimal error probability can be characterized by
\begin{equation}
\nu(s,s)=\min_{(p_{Y|X,h_{0}},p_{Y|X,h_{1}})\in\mathcal{P}_{\textnormal{E}}(s,s)}\left\{\textnormal{C}(p_{Y|h_{0}},p_{Y|h_{1}})\right\}.
\label{equation32}
\end{equation}
Similarly, the following theorem shows that the supply alphabet $\mathcal{Y}$ can be limited to the demand alphabet $\mathcal{X}$ without loss of optimality for the numerical evaluation of $\nu(s,s)$ in the considered smart meter privacy context.

\begin{theorem}
The energy supply alphabet can be limited to the energy demand alphabet under both hypotheses without loss of optimality for the evaluation of $\nu(s,s)$.
\label{theorem5.8}
\end{theorem}

\begin{IEEEproof}
Suppose that $\nu(s,s)=\textnormal{C}(p_{Y|h_{0}}^{*},p_{Y|h_{1}}^{*})$ is achieved by $p_{Y|X,h_{0}}^{*}$ and $p_{Y|X,h_{1}}^{*}$. We use the same quantization operation in the proof of Theorem \ref{theorem3} to map $y$ to $\hat{y}$. Therefore, $p_{\hat{Y}|X,h_{0}}$ and $p_{\hat{Y}|X,h_{1}}$ satisfy the constraints in the definition of $\mathcal{P}_{\textnormal{E}}(s,s)$. From the optimality in the definition of $\nu(s,s)$, it follows that
\begin{equation*}
\nu(s,s)=\textnormal{C}(p_{Y|h_{0}}^{*},p_{Y|h_{1}}^{*})\leq\textnormal{C}(p_{\hat{Y}|h_{0}},p_{\hat{Y}|h_{1}}).
\end{equation*}
In addition, from the data processing inequality of R\'{e}nyi divergence \cite[Theorem 9]{tim2014}, we have
\begin{equation*}
\begin{aligned}
&\textnormal{C}(p_{Y|h_{0}}^{*},p_{Y|h_{1}}^{*})\\
=&\max_{0\leq\tau\leq1}\left\{\textnormal{C}_{\tau}(p_{Y|h_{0}}^{*},p_{Y|h_{1}}^{*})\right\}\\
=&\max\{\max_{0\leq\tau<1}\{(1-\tau)\textnormal{D}_{\tau}(p_{Y|h_{0}}^{*}||p_{Y|h_{1}}^{*})\},\textnormal{D}_{0}(p_{Y|h_{1}}^{*}||p_{Y|h_{0}}^{*})\}\\
\geq&\max\{\max_{0\leq\tau<1}\{(1-\tau)\textnormal{D}_{\tau}(p_{\hat{Y}|h_{0}}||p_{\hat{Y}|h_{1}})\},\textnormal{D}_{0}(p_{\hat{Y}|h_{1}}||p_{\hat{Y}|h_{0}})\}\\
=&\max_{0\leq\tau\leq1}\left\{\textnormal{C}_{\tau}(p_{\hat{Y}|h_{0}},p_{\hat{Y}|h_{1}})\right\}\\
=&\textnormal{C}(p_{\hat{Y}|h_{0}},p_{\hat{Y}|h_{1}}).
\end{aligned}
\end{equation*}
Therefore,
\begin{equation*}
\nu(s,s)=\textnormal{C}(p_{\hat{Y}|h_{0}},p_{\hat{Y}|h_{1}}),
\end{equation*}
and the energy supply alphabet under both hypotheses can be constrained to $\mathcal{X}$ without loss of optimality.
\end{IEEEproof}

We also highlight here the connections and differences of our result with other smart meter privacy literature. While Kullback-Leibler divergence can be considered as yet another statistical similarity measure, our formulation here provides an operational meaning to its use as the privacy measure. This is in contrast to some of the other privacy measures considered in the literature, such as the mutual information used in \cite{varodayan2011,tan2013,jesus2015}, which are mainly proposed as distances between the energy demand data and the smart meter reading. A related work is \cite{farokhi2018} in the context of smart meter privacy, which uses the Fisher information as a privacy measure. The Fisher information is an approximation of the Kullback-Leibler divergence between two similar probability distributions \cite{jeffreys1946}; therefore, the two privacy measures are closely related. But the authors in \cite{farokhi2018} focus on a parameter estimation problem, rather than a classification problem in our work, and exploit an energy storage device to provide privacy. It is difficult to compare our result with \cite{varodayan2011,tan2013,jesus2015,farokhi2018} because of their different problem settings, e.g., there is only one energy demand profile in \cite{varodayan2011,tan2013,jesus2015,farokhi2018} while there are two energy demand profiles in our work. The connection between differential privacy and the minimum mutual information leakage problem has been revealed in \cite{wang2016}. The differential privacy model in \cite{acs2011,zhao2014} considers the aggregation of meter readings rather than a sequence of meter readings in our work; and furthermore, the constraint of non-negative renewable energy supply in our work cannot provide the Laplace noise often used in the differential privacy model.

\subsection{Binary Demand Example}
\begin{figure}
\centering
\includegraphics[scale=0.45]{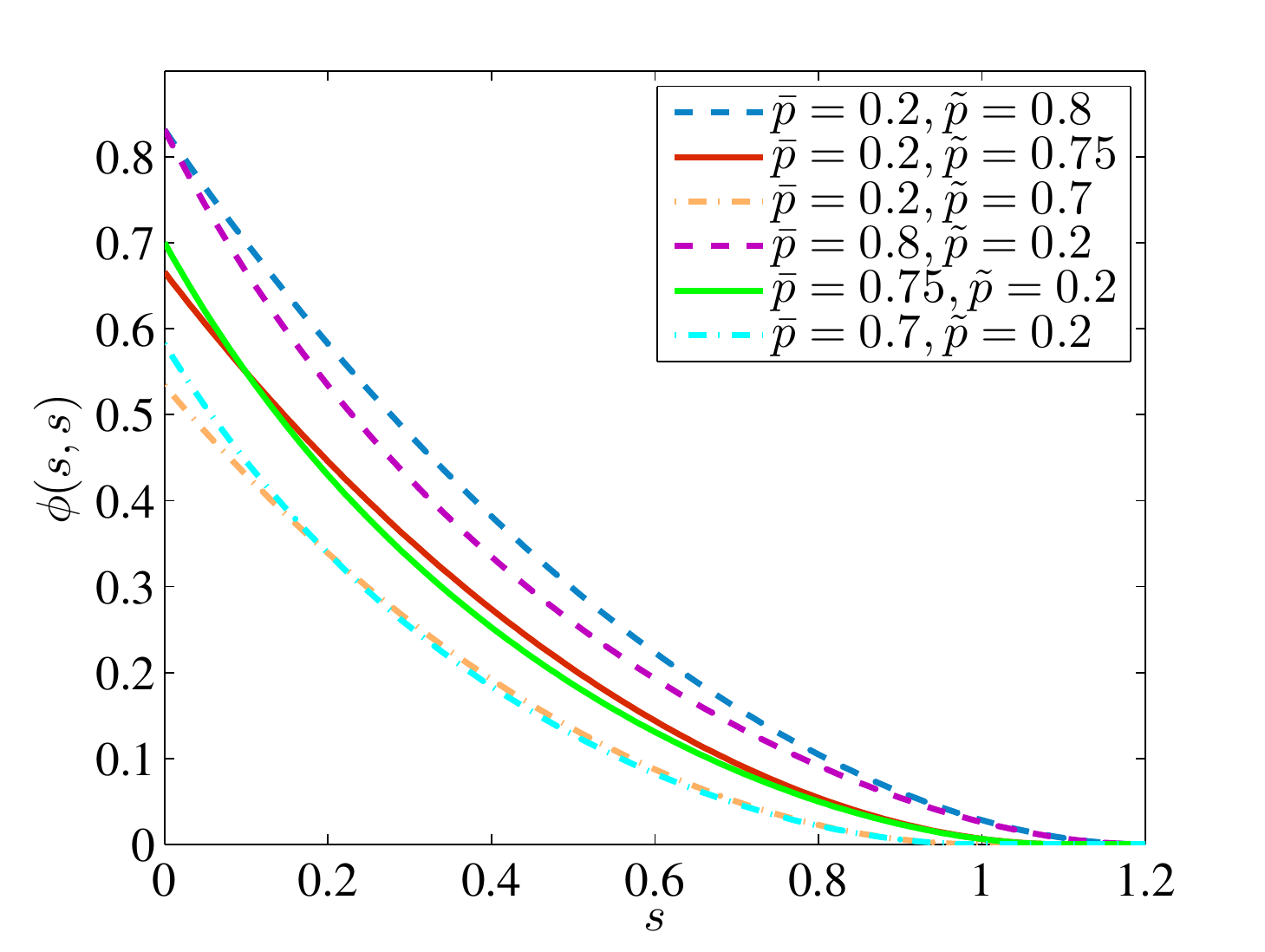}
\caption{Asymptotic privacy performance $\phi(s,s)$ for a binary demand model with different values of $\bar{p}$ and $\tilde{p}$.}
\label{figure2}
\end{figure}

\begin{figure}
\centering
\includegraphics[scale=0.45]{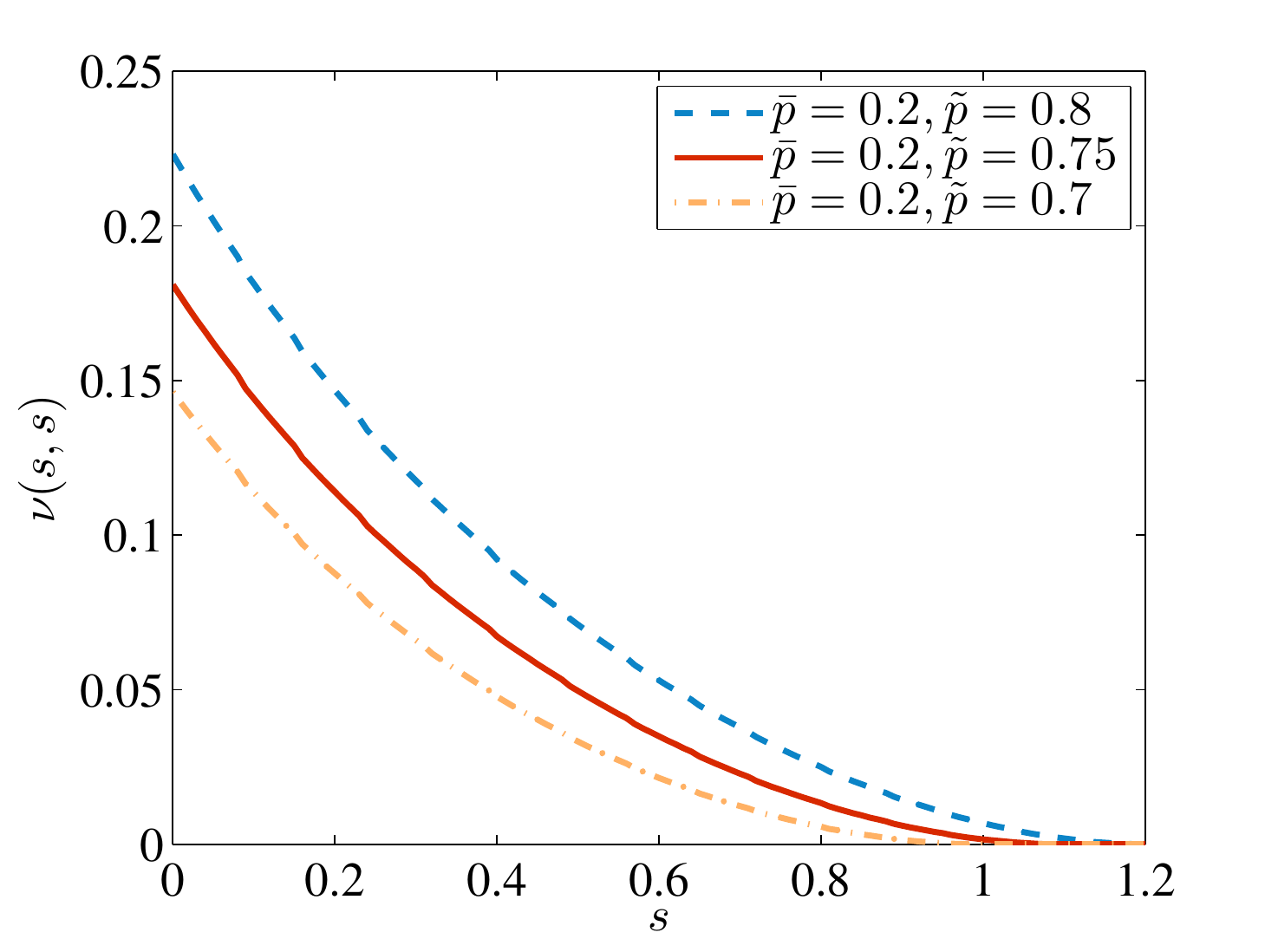}
\caption{Asymptotic privacy performance $\nu(s,s)$ for a binary demand model with different values of $\bar{p}$ and $\tilde{p}$.}
\label{figure5.3}
\end{figure}
We first present a simple example with binary energy demand alphabet\footnote{Certain appliances have binary energy demands, e.g., a fridge with ``on" and ``sleep" modes.} $\mathcal{X}=\{0,2\}$ for the purpose to numerically illustrate the main results in this paper. Based on Theorem \ref{theorem3} and Theorem \ref{theorem5.8}, it is sufficient to consider a binary supply alphabet $\mathcal{Y}=\{0,2\}$, as well. Denote $p_{X|h_{0}}(0)$ by $\bar{p}$ and $p_{X|h_{1}}(0)$ by $\tilde{p}$.

Under the adversarial Neyman-Pearson test setting and using the optimal memoryless hypothesis-aware policy, the asymptotic privacy performance $\phi(s,s)$ is shown in Fig. \ref{figure2} for different values of $\bar{p}$ and $\tilde{p}$. Confirming the claim in Lemma \ref{lemma2}, it can be observed that $\phi(s,s)$ is convex and non-increasing. When $s=0$, $X_{i}=Y_{i}$ for all $i\in\mathbb{Z}_{+}$ under both hypotheses, and $\phi(0,0)=\textnormal{D}(p_{X|h_{0}}||p_{X|h_{1}})$. Intuitively, it is more difficult for the AD to identify the hypotheses when they lead to more similar energy demand profiles. It can be observed in Fig. \ref{figure2} that $\phi(s,s)$ decreases as $\tilde{p}$ (resp. $\bar{p}$) gets closer to the fixed $\bar{p}$ (resp. $\tilde{p}$). Another interesting observation is that $\phi(s,s)$ curves for different settings of energy demand statistics $(\bar{p},\tilde{p})$ might intersect. For instance, to achieve the privacy performance of $0.6$, a lower renewable energy generation rate is required for $(\bar{p}=0.2,\tilde{p}=0.75)$ than that for $(\bar{p}=0.75,\tilde{p}=0.2)$; while to achieve the privacy performance of $0.3$, a higher renewable energy generation rate is required for $(\bar{p}=0.2,\tilde{p}=0.75)$ than that for $(\bar{p}=0.75,\tilde{p}=0.2)$.

Under the adversarial Bayesian hypothesis test setting and using the optimal memoryless hypothesis-aware policy, the asymptotic privacy performance $\nu(s,s)$ is shown in Fig. \ref{figure5.3} for different values of $\bar{p}$ and $\tilde{p}$. Confirming the claim in Lemma \ref{lemma8}, the asymptotic privacy performance $\nu(s,s)$ is a convex and non-increasing function of $s$. From the same argument that more similar energy demand profiles make the AD more difficult to identify the hypotheses, it follows that $\nu(s,s)$ decreases as $\tilde{p}$ gets closer to the fixed $\bar{p}$. Note that the ``opposite'' settings, $(\bar{p}=0.8,\tilde{p}=0.2)$, $(\bar{p}=0.75,\tilde{p}=0.2)$, and $(\bar{p}=0.7,\tilde{p}=0.2)$, are not presented here since they lead to the same privacy performances as presented in the figure.

From the numerical results, the renewable energy generation rate that can guarantee a certain privacy performance can be determined.

\begin{table}
\caption{Operation modes under each hypothesis.}
\label{table1}
\begin{center}
\begin{tabular}{|l|c|c|c|c|}
\hline
\diagbox{$h$}{$p_{X|h}$}{$x$ [W]}&$0$&$200$&$500$&$1200$\\
\hline
$h_{0}$ (type A)&$0.2528$&$0.3676$&$0$&$0.3796$\\
\hline
$h_{1}$ (type B)&$0.1599$&$0.0579$&$0.2318$&$0.5504$\\
\hline
\end{tabular}
\end{center}
\end{table}

\subsection{Numerical Experiment}
\begin{figure}[t]
\centering
\includegraphics[scale=0.325]{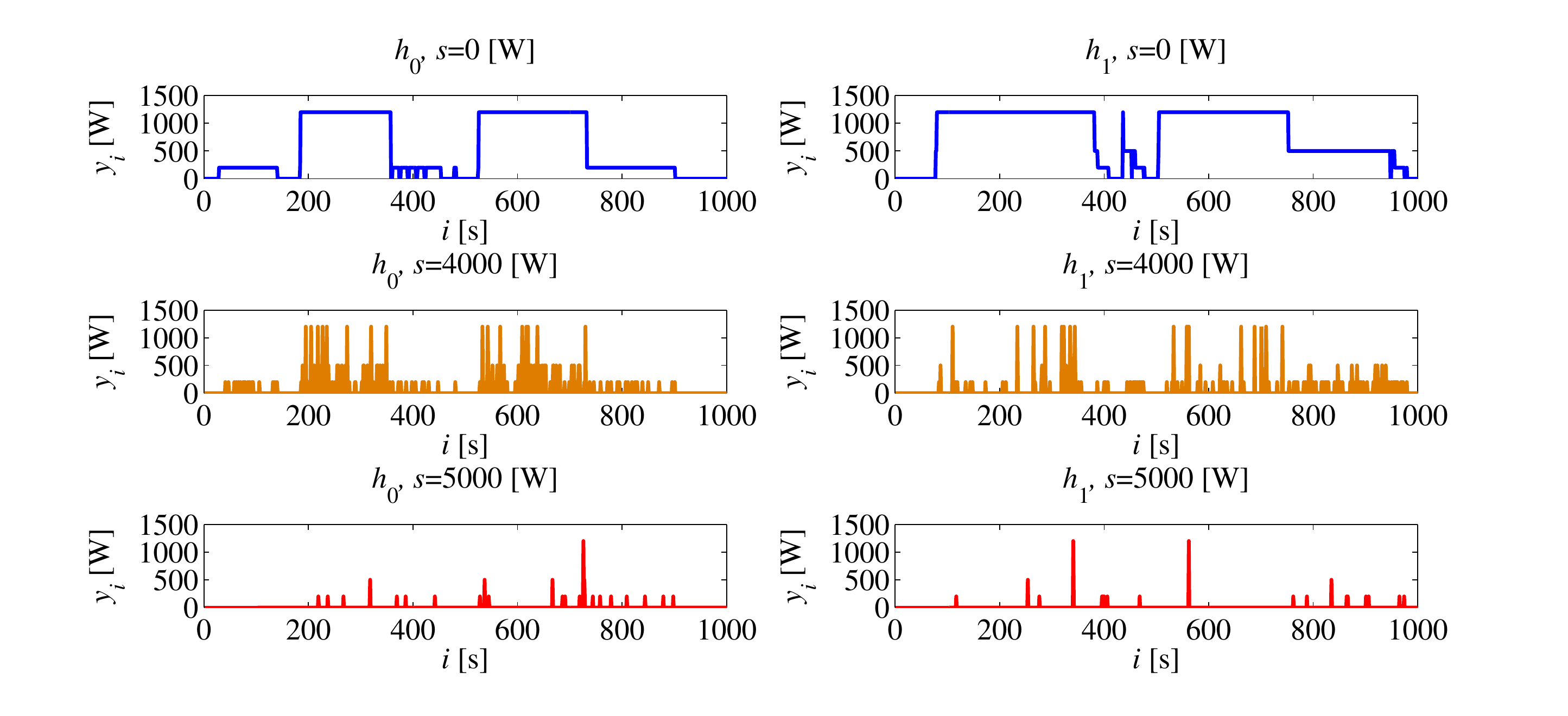}
\caption{Energy supplies under each hypothesis for different renewable energy generation rates when the optimal memoryless hypothesis-aware policy is used.}
\label{figure6}
\end{figure}

Here we present a numerical experiment with energy data from the REDD dataset \cite{kolter2011} to illustrate the practical value of our theoretic results. We consider a kitchen with a dishwasher, which can be type A ($h_{0}$) or type B ($h_{1}$). From the energy data, we can identify four operation modes of a dishwasher. Table \ref{table1} shows the operation modes and the corresponding statistics obtained through training under the assumption of i.i.d. energy demands. We consider three renewable energy generation rates: $s=0$, $4000$, $5000$ [W]. Given a value of $s$ and under the adversarial Neyman-Pearson test setting, the optimal memoryless hypothesis-aware policy is implemented on the real energy demand data under each hypothesis. The resulting energy supplies under each hypothesis are shown in Fig. \ref{figure6}. When $s=0$ [W], the energy supplies exactly follow the energy demands and it is easy for the AD to identify the hypothesis from the energy supply data. When $s=4000$ [W], the optimal policy enforces that every operation mode under hypothesis $h_{0}$ and the same operation mode under hypothesis $h_{1}$ are statistically similar. When $s=5000$ [W], most energy demands are satisfied from the RES, and it becomes very difficult for the AD to identify the hypothesis from the energy supply data.

\section{Conclusions}
\label{section6}
We have proposed novel formulations for privacy problem as adversarial hypothesis tests, where the fully-informed adversary makes an optimal hypothesis test on the privacy based on the adversarial observations; and the management unit manipulates the original data sequence into adversarial observations under an average distortion constraint to degrade the adversarial hypothesis testing accuracy. The privacy performance is evaluated by an asymptotic error exponent of the fully-informed adversary.

In the adversarial Neyman-Pearson hypothesis test setting, an asymptotically optimal privacy performance is shown to be characterized by a Kullback-Leibler divergence rate. Focusing on the worst case scenario, in which the type I probability of error upper bound approaches one, the asymptotic exponent of the maximum minimal type II probability of error achieved by the optimal memoryless hypothesis-aware policy is characterized by a single-letter Kullback-Leibler divergence; and it is also shown that the optimal memoryless hypothesis-aware policy cannot asymptotically outperform the optimal hypothesis-unaware policy with memory, since the management unit can learn the true hypothesis with an arbitrarily small probability of error from a sufficiently long sequence of original data. In the adversarial Bayesian hypothesis test setting, the informed adversary is assumed to have access to the prior knowledge on the hypotheses, and the privacy leakage is measured by the minimal error probability of the adversary. Asymptotic results are derived on the optimal privacy performance similar to the adversarial Neyman-Pearson hypothesis test setting by substituting Kullback-Leibler divergence with Chernoff information. These theoretic results can be directly applied to the smart meter privacy problem, where the energy supply alphabet can be constrained to the energy demand alphabet without loss of optimality for the evaluation of the single-letter privacy performances in both adversarial hypothesis test settings.

The asymptotic privacy performances derived in this work provide fundamental limits on any practical management policy and they therefore serve as references/benchmarks useful for the performance assessment of any practical policy. The optimal memoryless hypothesis-aware policy is an i.i.d. policy, which can be easily designed and implemented in practice.

\appendix
\subsection{Proof of Theorem \ref{theorem2}}
\begin{IEEEproof}
Let $o(n)\triangleq\lfloor\log n\rfloor$, $c(n)\triangleq o(n)+1$, $q(n)\triangleq n-o(n)$, and $d_{\max}\triangleq\max_{x\in\mathcal{X},y\in\mathcal{Y}}d(x,y)$. We choose any $\delta\in(0,s)$, $\omega\in(0,s)$, and type I probability of error upper bound $\varepsilon'\in(\max\{0,1-\min\{\textnormal{D}(p_{X|h_{0}}||p_{X|h_{1}}),\frac{\delta}{d_{\max}}\}\},1)$. Then set $\xi=1-\varepsilon'$ and $\psi=\delta-d_{\max}\cdot\xi$. It can be verified that $0<\xi<\textnormal{D}(p_{X|h_{0}}||p_{X|h_{1}})$ and $0<\psi\leq\delta$. We use these parameters to construct a hypothesis-unaware policy with memory, $\rho^{n}_{\textnormal{p}}$, over an $n$-slot time horizon, which consists of two successive phases.

$o(n)${\it -slot learning phase}. The goal of the MU is to learn the true hypothesis at the end of the first phase. To prevent privacy leakage during the learning phase, identical instantaneous policies are used at all time slots as:
\begin{equation*}
y_{i}=\rho_{i}(x_{i})=\min\mathcal{Y},\textnormal{ }\forall i\leq o(n),\textnormal{ }\forall x_{i}\in\mathcal{X}.
\end{equation*}
Based on the observations of original data $x^{o(n)}$, the MU makes a decision $\hat{H}$:
\begin{equation*}
\hat{H}=\left\{
\begin{aligned}
&h_{0},\textnormal{ if }x^{o(n)}\in\mathcal{A}_{\xi}^{o(n)}(p_{X|h_{0}}||p_{X|h_{1}})\\
&h_{1},\textnormal{ otherwise}
\end{aligned}
\right.,
\end{equation*}
where $\mathcal{A}_{\xi}^{o(n)}(p_{X|h_{0}}||p_{X|h_{1}})$ denotes a relative entropy typical set as defined in \cite{cover2006}, and any sequence $x^{o(n)}\in\mathcal{A}_{\xi}^{o(n)}(p_{X|h_{0}}||p_{X|h_{1}})$ satisfies
\begin{equation*}
\left|\frac{1}{o(n)}\log\frac{p_{X^{o(n)}|h_{0}}\left(x^{o(n)}\right)}{p_{X^{o(n)}|h_{1}}\left(x^{o(n)}\right)}-\textnormal{D}(p_{X|h_{0}}||p_{X|h_{1}})\right|\leq\xi.
\end{equation*}
The MU can make a wrong decision. Let $p_{\bar{e}}=1-p_{X^{o(n)}|h_{0}}(\mathcal{A}_{\xi}^{o(n)}(p_{X|h_{0}}||p_{X|h_{1}}))$ denote the type I probability of error and $p_{\tilde{e}}=p_{X^{o(n)}|h_{1}}(\mathcal{A}_{\xi}^{o(n)}(p_{X|h_{0}}||p_{X|h_{1}}))$ denote the type II probability of error for the MU.

$q(n)${\it -slot privacy-preserving phase}. Depending on the decision $\hat{H}$ in the learning phase, identical instantaneous policies are used at all slots of the second phase as follows.\newline
If $\hat{H}=h_{0}$ (resp. $\hat{H}=h_{1}$) and for all $i\in\{c(n),\dots,n\}$, $\rho_{i}:p_{Y_{i}|X_{i}}$ corresponds to the optimizer $p_{Y|X,h_{0}}^{*}$ (resp. $p_{Y|X,h_{1}}^{*}$) of $\phi(s-\delta,s-\omega)$.

Next, we check the average distortion constraint. Under hypothesis $h_{0}$ (resp. $h_{1}$), if a correct decision is made at the end of the learning phase, the instantaneous management policies used in the privacy-preserving phase guarantee that the single-slot average distortion upper bound $s-\delta$ (resp. $s-\omega$) is satisfied at all time slots of this phase; otherwise, bounded violation of the single-slot average distortion constraint might happen at all time slots of the privacy-preserving phase. When $n$ is sufficiently large, we have
\begin{equation*}
\begin{aligned}
&\textnormal{E}\left[\left.\frac{1}{n}\sum_{i=1}^{n}d(X_{i},Y_{i})\right|h_{0}\right]\\
\leq&\frac{o(n)}{n}d_{\max}+\textnormal{E}\left[\left.\frac{1}{n}\sum_{i=c(n)}^{n}d(X_{i},Y_{i})\right|h_{0}\right]\\
\overset{(a)}{\leq}&\frac{o(n)}{n}d_{\max}+\frac{q(n)}{n}(s-\delta+d_{\max}\cdot\xi)\\
\leq&s-\delta+d_{\max}\cdot\xi+\frac{o(n)}{n}d_{\max}\\
=&s-(\psi-\frac{o(n)}{n}d_{\max})\\
\overset{(b)}{\leq}&s,
\end{aligned}
\end{equation*}
and
\begin{equation*}
\begin{aligned}
&\textnormal{E}\left[\left.\frac{1}{n}\sum_{i=1}^{n}d(X_{i},Y_{i})\right|h_{1}\right]\\
\leq&\frac{o(n)}{n}d_{\max}+\textnormal{E}\left[\left.\frac{1}{n}\sum_{i=c(n)}^{n}d(X_{i},Y_{i})\right|h_{1}\right]\\
\leq&\frac{o(n)}{n}d_{\max}+\frac{q(n)}{n}(s-\omega+d_{\max}\cdot p_{\tilde{e}})\\
\overset{(c)}{\leq}&s+\underbrace{d_{\max}\cdot e^{-o(n)\cdot(\textnormal{D}(p_{X|h_{0}}||p_{X|h_{1}})-\xi)}+\frac{o(n)}{n}d_{\max}-\omega}_{\Delta(n)}\\
\overset{(d)}{\leq}&s,
\end{aligned}
\end{equation*}
where $(a)$ follows since $p_{\bar{e}}<\xi$ when $o(n)$ is sufficiently large \cite[Theorem 11.8.2]{cover2006}; $(b)$ follows since $\psi-\frac{o(n)}{n}d_{\max}\geq0$ when $n$ is sufficiently large; $(c)$ follows since $p_{\tilde{e}}<e^{-o(n)\cdot(\textnormal{D}(p_{X|h_{0}}||p_{X|h_{1}})-\xi)}$ \cite[Theorem 11.8.2]{cover2006}; and $(d)$ follows since $\Delta(n)\leq0$ when $n$ is sufficiently large.

Therefore, $\rho^{n}_{\textnormal{p}}$ is a management policy which satisfies the average distortion constraint in (\ref{equation2}) when $n$ is sufficiently large. Then, we have
\begin{equation}
\begin{aligned}
\liminf_{n\to\infty}\frac{1}{q(n)}\log\frac{1}{\beta(n,\varepsilon',\rho^{n}_{\textnormal{p}})}\geq&\liminf_{n\to\infty}\frac{1}{n}\log\frac{1}{\beta(n,\varepsilon',\rho^{n}_{\textnormal{p}})}\\
\geq&\liminf_{n\to\infty}\frac{1}{n}\log\frac{1}{\beta_{\textnormal{M}}(n,\varepsilon',s)},
\end{aligned}
\label{equation25}
\end{equation}
where the first inequality follows from $q(n)\leq n$; and the second inequality follows since the constructed management policy $\rho^{n}_{\textnormal{p}}$ is not necessarily optimal for all sufficiently large $n$.

From the point of view of the informed AD, the observations $y^{o(n)}$ in the learning phase do not reveal any information about the hypothesis. Therefore, the strategy of the AD only depends on the observation statistics in the privacy-preserving phase $p_{Y_{c(n)}^{n}|h_{0}}$ and $p_{Y_{c(n)}^{n}|h_{1}}$. Then, the term $\frac{1}{q(n)}\log\frac{1}{\beta(n,\varepsilon',\rho^{n}_{\textnormal{p}})}$ in (\ref{equation25}) represents the exponent of the minimal type II probability of error in the privacy-preserving phase. With the management policy $\rho_{\textnormal{p}}^{n}$ specified above, the sequence of adversarial observations in the privacy-preserving phase is a mixture of the i.i.d. sequences $Y_{c(n)}^{n}|\hat{H}=h_{0},H=h_{0}$ and $Y_{c(n)}^{n}|\hat{H}=h_{1},H=h_{0}$ under hypothesis $h_{0}$, or a mixture of the i.i.d. sequences $Y_{c(n)}^{n}|\hat{H}=h_{1},H=h_{1}$ and $Y_{c(n)}^{n}|\hat{H}=h_{0},H=h_{1}$ under hypothesis $h_{1}$. The probability distributions are
\begin{equation*}
\begin{aligned}
p_{Y_{c(n)}^{n}|h_{0}}\left(y_{c(n)}^{n}\right)=&(1-p_{\bar{e}})\cdot p_{Y_{c(n)}^{n}|\hat{H}=h_{0},H=h_{0}}\left(y_{c(n)}^{n}\right)\\
&+p_{\bar{e}}\cdot p_{Y_{c(n)}^{n}|\hat{H}=h_{1},H=h_{0}}\left(y_{c(n)}^{n}\right),\\
p_{Y_{c(n)}^{n}|h_{1}}\left(y_{c(n)}^{n}\right)=&(1-p_{\tilde{e}})\cdot p_{Y_{c(n)}^{n}|\hat{H}=h_{1},H=h_{1}}\left(y_{c(n)}^{n}\right)\\
&+p_{\tilde{e}}\cdot p_{Y_{c(n)}^{n}|\hat{H}=h_{0},H=h_{1}}\left(y_{c(n)}^{n}\right).
\end{aligned}
\end{equation*}
We define
\begin{equation*}
\mathcal{B}(R,n)\triangleq\left\{y_{c(n)}^{n}\left|\frac{1}{q(n)}\log\frac{p_{Y_{c(n)}^{n}|h_{0}}\left(y_{c(n)}^{n}\right)}{p_{Y_{c(n)}^{n}|h_{1}}\left(y_{c(n)}^{n}\right)}\right.\leq R\right\},
\end{equation*}
and
\begin{equation*}
K(R)\triangleq\limsup_{n\to\infty}p_{Y_{c(n)}^{n}|h_{0}}\left(\mathcal{B}(R,n)\right).
\end{equation*}
Based on the information-spectrum results \cite[Theorem 1]{chen1996}, \cite[Theorem 4.2.1]{han2003}, we have
\begin{equation}
\sup\{R|K(R)\leq\varepsilon'\}\geq\liminf_{n\to\infty}\frac{1}{q(n)}\log\frac{1}{\beta(n,\varepsilon',\rho^{n}_{\textnormal{p}})}.
\label{equation26}
\end{equation}

In the asymptotic regime as $n\to\infty$, $Y_{c(n)}^{n}|h_{1}$ reduces to the i.i.d. sequence $Y_{c(n)}^{n}|\hat{H}=h_{1},H=h_{1}$ since $\lim_{n\to\infty}p_{\tilde{e}}\leq\lim_{n\to\infty}e^{-o(n)\cdot(\textnormal{D}(p_{X|h_{0}}||p_{X|h_{1}})-\xi)}=0$. Based on the case study \cite[Example 4.2.1]{han2003}, the upper bound $\sup\{R|K(R)\leq\varepsilon'\}$ is characterized by the Kullback-Leibler divergences $\textnormal{D}_{1}=\textnormal{D}(p_{Y|\hat{H}=h_{0},H=h_{0}}||p_{Y|\hat{H}=h_{1},H=h_{1}})$ and $\textnormal{D}_{2}=\textnormal{D}(p_{Y|\hat{H}=h_{1},H=h_{0}}||p_{Y|\hat{H}=h_{1},H=h_{1}})$ as follows.\newline
If $\textnormal{D}_{1}\geq\textnormal{D}_{2}$, we have
\begin{equation*}
\textnormal{D}(p_{Y|\hat{H}=h_{0},H=h_{0}}||p_{Y|\hat{H}=h_{1},H=h_{1}})\geq\sup\{R|K(R)\leq\varepsilon'\};
\end{equation*}
otherwise, since $1-p_{\bar{e}}>1-\xi=\varepsilon'$ as $n\to\infty$, we have
\begin{equation*}
\textnormal{D}(p_{Y|\hat{H}=h_{0},H=h_{0}}||p_{Y|\hat{H}=h_{1},H=h_{1}})=\sup\{R|K(R)\leq\varepsilon'\}.
\end{equation*}
Then, it follows that
\begin{equation}
\begin{aligned}
\phi(s-\delta,s-\omega)=&\textnormal{D}(p_{Y|\hat{H}=h_{0},H=h_{0}}||p_{Y|\hat{H}=h_{1},H=h_{1}})\\
\geq&\sup\{R|K(R)\leq\varepsilon'\}.
\end{aligned}
\label{equation27}
\end{equation}

The inequalities (\ref{equation25}), (\ref{equation26}), and (\ref{equation27}) jointly lead to
\begin{equation}
\phi(s-\delta,s-\omega)\geq\liminf_{n\to\infty}\frac{1}{n}\log\frac{1}{\beta_{\textnormal{M}}(n,\varepsilon',s)}.
\label{equation29}
\end{equation}

Given $\delta$, $\omega\in(0,s)$, the inequality (\ref{equation29}) holds for all type I probability of error upper bounds $\varepsilon'\in(\max{\{0,1-\min\{\textnormal{D}(p_{X|h_{0}}||p_{X|h_{1}}),\frac{\delta}{d_{\max}}\}\}},1)$. Therefore,
\begin{equation}
\phi(s-\delta,s-\omega)\geq\lim_{\varepsilon\to1}\liminf_{n\to\infty}\frac{1}{n}\log\frac{1}{\beta_{\textnormal{M}}(n,\varepsilon,s)}=\theta_{\textnormal{M}}(s).
\label{equation28}
\end{equation}
Since the inequality (\ref{equation28}) holds for all $\delta$, $\omega\in(0,s)$, we further have
\begin{equation*}
\phi(s,s)=\inf_{\delta,\omega\in(0,s)}\{\phi(s-\delta,s-\omega)\}\geq\theta_{\textnormal{M}}(s),
\end{equation*}
where the equality follows from the non-increasing and continuous properties of $\phi(\bar{s},\tilde{s})$.
\end{IEEEproof}

\subsection{Proof of Theorem \ref{theorem7}}
\begin{IEEEproof}
Let $o(n)\triangleq\lfloor\log n\rfloor$, $c(n)\triangleq o(n)+1$, $q(n)\triangleq n-o(n)$, $d_{\max}\triangleq\max_{x\in\mathcal{X},y\in\mathcal{Y}}d(x,y)$. We choose any $\delta\in(0,s)$, $\omega\in(0,s)$, and $\xi\in\left(0,\min\left\{\textnormal{D}(p_{X|h_{0}}||p_{X|h_{1}}),\frac{\delta}{d_{\max}}\right\}\right)$. Then set $\psi=\delta-d_{\max}\cdot\xi$. It can be verified that $0<\psi<\delta$. We use these parameters to construct a hypothesis-unaware management policy with memory, $\rho^{n}_{\textnormal{q}}$, over an $n$-slot time horizon, which similarly has two phases as $\rho^{n}_{\textnormal{p}}$.

$o(n)${\it -slot learning phase}. This phase is the same as the learning phase of $\rho_{\textnormal{p}}^{n}$: Identical instantaneous privacy-unaware policies are independently used at all slots and always output an adversarial observation sequence of $\min\mathcal{Y}$ regardless of the sequence of original data; and the learning decision at the end of this phase is $\hat{H}=0$ if the sequence $x^{o(n)}$ is in the relative entropy typical set $\mathcal{A}_{\xi}^{o(n)}\left(p_{X|h_{0}}||p_{X|h_{1}}\right)$ or is $\hat{H}=1$ otherwise. Again, we denote the type I probability of learning error by $p_{\bar{e}}$ and the type II probability of learning error by $p_{\tilde{e}}$.

$q(n)${\it -slot privacy-preserving phase}. Depending on the decision $\hat{H}$ in the learning phase, identical instantaneous management policies are used at all slots of the second phase as follows.\newline
If $\hat{H}=h_{0}$ (resp. $\hat{H}=h_{1}$) and for all $i\in\{c(n),\dots,n\}$, $\rho_{i}:p_{Y_{i}|X_{i}}$ corresponds to the optimizer $p_{Y|X,h_{0}}^{*}$ (resp. $p_{Y|X,h_{1}}^{*}$) of $\nu(s-\delta,s-\omega)$.

It follows from the same analysis on $\rho_{\textnormal{p}}^{n}$ that the policy $\rho^{n}_{\textnormal{q}}$ satisfies the average distortion constraint in (\ref{equation2}) when $n$ is sufficiently large. Similar to (\ref{equation25}), we have
\begin{equation}
\begin{aligned}
\liminf_{n\to\infty}\frac{1}{q(n)}\log\frac{1}{\alpha(n,\rho^{n}_{\textnormal{q}})}\geq&\liminf_{n\to\infty}\frac{1}{n}\log\frac{1}{\alpha(n,\rho^{n}_{\textnormal{q}})}\\
\geq&\lim_{n\to\infty}\frac{1}{n}\log\frac{1}{\alpha_{\textnormal{M}}(n,s)}=\mu_{\textnormal{M}}(s).
\end{aligned}
\label{equation88}
\end{equation}

Since $y^{o(n)}$ in the learning phase is a deterministic sequence of $\min\mathcal{Y}$, the strategy of the AD only depends on the observation statistics in the privacy-preserving phase $p_{Y_{c(n)}^{n}|h_{0}}$ and $p_{Y_{c(n)}^{n}|h_{1}}$. Then, the term $\frac{1}{q(n)}\log\frac{1}{\alpha(n,\rho^{n}_{\textnormal{q}})}$ in (\ref{equation88}) represents the exponent of the minimal error probability in the privacy-preserving phase. With the management policy $\rho_{\textnormal{q}}^{n}$ specified above, the sequence of adversarial observations in the privacy-preserving phase is a mixture of the i.i.d. sequences $Y_{c(n)}^{n}|\hat{H}=h_{0},H=h_{0}$ with a probability $1-p_{\bar{e}}$ and $Y_{c(n)}^{n}|\hat{H}=h_{1},H=h_{0}$ with a probability $p_{\bar{e}}$ under hypothesis $h_{0}$, or a mixture of the i.i.d. sequences $Y_{c(n)}^{n}|\hat{H}=h_{1},H=h_{1}$ with a probability $1-p_{\tilde{e}}$ and $Y_{c(n)}^{n}|\hat{H}=h_{0},H=h_{1}$ with a probability $p_{\tilde{e}}$ under hypothesis $h_{1}$. In the asymptotic regime as $n\to\infty$, $Y_{c(n)}^{n}|h_{1}$ can be seen as the i.i.d. sequence $Y_{c(n)}^{n}|\hat{H}=h_{1},H=h_{1}$ since $\lim_{n\to\infty}p_{\tilde{e}}\leq\lim_{n\to\infty}e^{-o(n)\cdot(\textnormal{D}(p_{X|h_{0}}||p_{X|h_{1}})-\xi)}=0$. Then, we have
\begin{equation}
\begin{aligned}
&\liminf_{n\to\infty}\frac{1}{q(n)}\log\frac{1}{\alpha(n,\rho^{n}_{\textnormal{q}})}\\
=&\textnormal{C}(p_{Y|\hat{H}=h_{0},H=h_{0}},p_{Y|\hat{H}=h_{1},H=h_{1}})+\Delta(\xi)\\
=&\nu(s-\delta,s-\omega)+\Delta(\xi),
\end{aligned}
\label{equation89}
\end{equation}
where $\nu(s-\delta,s-\omega)$ corresponds to the asymptotic exponent of the minimal error probability in the privacy-preserving phase if $p_{\bar{e}}=0$; and the term $\Delta(\xi)$ denotes the impact of the i.i.d. sequence $Y_{c(n)}^{n}|\hat{H}=h_{1},H=h_{0}$ with a probability $p_{\bar{e}}$ under hypothesis $h_{0}$. The inequalities (\ref{equation88}) and (\ref{equation89}) jointly lead to
\begin{equation}
\mu_{\textnormal{M}}(s)\leq\nu(s-\delta,s-\omega)+\Delta(\xi).
\label{equation90}
\end{equation}

Since $\limsup_{n\to\infty}p_{\bar{e}}\leq\xi$, the term $\Delta(\xi)$ satisfies
\begin{equation}
\lim_{\xi\to0}\Delta(\xi)=0.
\label{equation91}
\end{equation}
Given $\delta$, $\omega\in(0,s)$, the inequality (\ref{equation90}) holds for all $\xi\in\left(0,\min\left\{\textnormal{D}(p_{X|h_{0}}||p_{X|h_{1}}),\frac{\delta}{d_{\max}}\right\}\right)$. Therefore,
\begin{equation}
\mu_{\textnormal{M}}(s)\leq\nu(s-\delta,s-\omega)+\lim_{\xi\to0}\Delta(\xi)=\nu(s-\delta,s-\omega).
\label{equation92}
\end{equation}
Since (\ref{equation92}) holds for all $\delta$, $\omega\in(0,s)$, we further have
\begin{equation}
\mu_{\textnormal{M}}(s)\leq\inf_{\delta,\omega\in(0,s)}\left\{\nu(s-\delta,s-\omega)\right\}=\nu(s,s),
\label{equation93}
\end{equation}
where the equality follows from the non-increasing and continuous properties of $\nu(\bar{s},\tilde{s})$.
\end{IEEEproof}

\bibliographystyle{IEEEtran}
\bibliography{IEEEabrv,References}

\end{document}